\documentclass[twocolumn,showpacs,preprintnumbers,nofootinbib,prd,
superscriptaddress,10pt]{revtex4-2}

\pdfoutput=1

\usepackage{amsmath,amssymb}
\usepackage{amsfonts}
\usepackage{mathtools}
\usepackage[normalem]{ulem}
\usepackage{textcomp}
\usepackage[bookmarks]{hyperref}
\usepackage{enumitem}
\usepackage{bm}
\usepackage{bbm}
\usepackage{afterpage}
\usepackage{graphicx}
\graphicspath{{img/}} 
\usepackage{tensor}
\usepackage{orcidlink}

\usepackage{tabularx}
\usepackage{arydshln}
\usepackage{makecell}
\usepackage{multirow}
\usepackage{booktabs}

\usepackage{layouts}
\usepackage[utf8]{inputenc}
\usepackage{blindtext}
\usepackage{graphicx}
\usepackage{siunitx}
\renewcommand{\d}[1]{\ensuremath{\operatorname{d}\!{#1}}}
\newcommand{\dvol}[2]{\ensuremath{\operatorname{d}^{#2}\!{#1}}}

\DeclareMathOperator*{\argmin}{arg\,min}

\newcommand{\scalar}[2]{\langle #1|#2 \rangle}

\newcommand{\rescalar}[2]{( #1 |#2 )}
\newcommand{\rescalarwide}[2]{\left( #1 \lvert #2 \right)}
\newcommand{\imscalar}[2]{[ #1|#2 ]}

\begin{document}

\begin{abstract}
	We introduce a novel method to generate a bank of gravitational-waveform templates of binary black hole (BBH) mergers for matched-filter searches in LIGO, Virgo and Kagra data.
	We derive a novel expression for the metric approximation to the distance between templates, which is suitable for precessing BBHs and/or systems with higher-order modes (HM) imprints and we use it to meaningfully define a template probability density across the parameter space.
	We employ a masked autoregressive normalizing flow model which can be conveniently trained to quickly reproduce the target probability distribution and sample templates from it.
	Thanks to the normalizing flow, our code takes a few {\it hours} to produce random template banks with millions of templates, making it particularly suitable for high-dimensional spaces, such as those associated to precession, eccentricity and/or HM.
	After validating the performance of our method, we generate a bank for precessing black holes and a bank for aligned-spin binaries with HMs: with only 5\% of the injections with fitting factor below the target of $0.97$, we show that both banks cover satisfactorily the space.
	Our publicly released code \texttt{mbank} will enable searches of high-dimensional regions of BBH signal space, hitherto unfeasible due to the prohibitive cost of bank generation.
	
\end{abstract}
	
 \title{Gravitational-wave template banks for novel compact binaries}
	\author{Stefano \surname{Schmidt} \orcidlink{0000-0002-8206-8089}} 
		\email{s.schmidt@uu.nl}
        \affiliation{Nikhef, Science Park 105, 1098 XG, Amsterdam, The Netherlands}
        \affiliation{Institute for Gravitational and Subatomic Physics (GRASP),
Utrecht University, Princetonplein 1, 3584 CC Utrecht, The Netherlands}

	\author{Bhooshan \surname{Gadre} \orcidlink{0000-0002-1534-9761}}
        \affiliation{Institute for Gravitational and Subatomic Physics (GRASP),
Utrecht University, Princetonplein 1, 3584 CC Utrecht, The Netherlands}

	\author{Sarah \surname{Caudill} \orcidlink{0000-0002-8927-6673}}
		\affiliation{Department of Physics, University of Massachusetts, Dartmouth, MA 02747, USA}
		\affiliation{Center for Scientific Computing and Visualization Research, University of Massachusetts, Dartmouth, MA 02747, USA}
	\maketitle
\section{Introduction}

As gravitational-wave (GW) astronomy enters a mature state, the accessible parameter space of binary black hole (BBH) mergers in LIGO \cite{LIGOScientific:2014pky} and Virgo \cite{VIRGO:2014yos} data continues to grow. Besides standard aligned-spin GW searches for stellar-mass BBH mergers \cite{GWTC-1,GWTC-2,GWTC-2.1, GWTC-3}, there are GW searches targeting the parameter space of sub-solar mass black holes (BH) \cite{SSM_O2, SSM_O3a, PhysRevD.106.023024, Nitz:2021mzz}, primordial BHs \cite{PBH}, eccentric binaries \cite{PhysRevD.102.043005, PhysRevD.104.104016, Nitz:2019spj, LIGOScientific:2019dag, Ramos-Buades:2020eju, Wang:2021qsu, Nitz:2021mzz} and intermediate-mass BHs (IMBH) \cite{IMBH_O2, IMBH_O3, Chandra:2022ixv}. Moreover, there is a growing interest in GW searches for more complex binaries, such as those with precession \cite{PhysRevD.89.024010, Harry:2017weg, PhysRevD.102.041302, Indik:2016qky, Harry:2016ijz, Fairhurst:2019vut, McIsaac:2023ijd} or higher-order mode (HMs) content~\cite{CalderonBustillo:2015lrt, Harry:2017weg, Chandra_hom, 2021PhRvD.103b4042M, Wadekar:2023kym}.

GW searches for signals from compact binary mergers traditionally utilize the method of matched-filtering with a template bank of model waveforms~\cite{Sathyaprakash:1991mt, Dhurandhar:1992mw, Owen:1998dk, Allen:2005fk, Babak:2006ty, Cokelaer:2007mv}.
An optimal template bank is composed of the smallest number of templates that guarantees that only a small fraction of signal-to-noise ratio from GW signals is missed due to the discreteness of the template bank \cite{Prix:2007ks}.

One widely used approach to bank generation - the {\it stochastic} method \cite{Harry:2009ea, PhysRevD.80.104014, Ajith:2012mn} - consists of randomly scattering templates in a defined parameter space with a rejection technique \cite{DalCanton:2017ala, Mukherjee:2018yra, Indik:2016qky, Lenon:2021zac}. A proposed template is included in the bank only if its distance (or {\it mismatch}) with all the proposed templates in the bank is larger than the user-defined threshold.
While this approach has proven to be very powerful, it does not scale well with (i) the number of templates and, most importantly, (ii) with the number of dimensions of the parameter space.

Handling a large number of templates can have a large impact on computing time and memory, because for every new proposal, a waveform needs to be generated and stored and many expensive match calculations need to be performed.
Furthermore, the sheer number of dimensions can have an even more catastrophic impact on the bank generation cost. Indeed, at every iteration the stochastic algorithm computes the distance between $N_\text{p}\sim r^D$ pairs of templates within a given radius $r$. It is clear how the number of match computations diverges for large dimensional spaces.

As the BBH searches grow in complexity due to the inclusion of more physical effects and hence more dimensions, the stochastic approach struggles to produce template banks in a feasible amount of time.
This poses the challenge of finding a viable alternative for template bank generation, which is able to deliver large banks in a high-dimensional parameter space, such as those associated with precession, eccentricity and HMs.

Revitalizing a pioneering line of research in bank generation \cite{owen_metric, Messenger:2008ta, Prix:2007ks, Brown:2012qf, Keppel:2013uma}, there has recently been increasing attention on {\it metric template placement} \cite{Roy:2017oul, Roulet:2019hzy, Coogan:2022qxs, Hanna:2022zpk}.
Such methods rely on approximating the distance (or mismatch) between two waveforms with a bilinear form, called metric.
Although the metric is only approximate, it allows for a faster template placing, which may overcome some of the major limitations of the standard stochastic placement algorithm.

Historically, the metric was first employed to place templates on a lattice \cite{owen_metric, Prix:2007ks, Cokelaer:2007kx}. However, constructing lattice-based template banks has proven to be challenging due to the difficulties in  obtaining coordinate transformation which avoids varying metric components.
To overcome such difficulties, a different metric placement method, called {\it random}, was introduced \cite{Messenger:2008ta} soon after.
Random template banks are designed to cover the region of interest with randomly sampled templates, without any control of the template spacing. 
Moreover, they are not designed to cover the whole space but only a large fraction $\eta<1$ of it (i.e., any point in space is covered with probability $\eta$).

The strength of the method is twofold: on the one hand, since no distance between templates is computed, the template placement is tremendously fast and memory efficient; on the other hand, by only covering a fraction of the space, the number of templates remains under control.
Moreover, the cost does not increase for an increasing number of dimensions.
While this may seem sub-optimal with respect to a lattice, in \cite{Messenger:2008ta, Allen:2022lqr, Allen:2021yuy} it is argued that for high-dimensional spaces, random template banks outperform even the best known lattice in terms of coverage (at a fixed number of templates), effectively beating the ``curse of dimensionality."

Generating a random template banks requires the ability to effectively sample templates ``uniformly" across the parameter space. Traditionally, due to the high dimensionality of the space, expensive sampling techniques, such as Markov Chain Montecarlo, must be used. This poses a serious limitation to the range of applicability of the method. Without a fast sampling method, the speed up promised by the new method is washed away by the cost of a large number of metric evaluations.

In this work, we address the challenges described above by covering high-dimensional spaces with random template banks.
As a first step, we derive a novel expression for the metric, which is suitable for generic precessing and/or HM waveforms. In doing so, we drop several symmetry assumptions that enters the standard metric computation. The metric is then expressed in terms of the gradients of the waveform.
Secondly, to enable a fast template sampling, we employ machine learning and train a {\it normalizing flow} model to efficiently sample templates from the parameter space.
While the first innovation delivers an accurate distribution for the templates throughout the space, the use of a normalizing flow allow us to generate random template banks in a few hours (including the training time).

The combination of a new metric expression and the normalizing flow model, applied to the random template placement algorithm, makes our method particularly well-suited for dealing with high-dimensional ($>$~4D) parameter spaces, such as those associated with precessing or eccentric searches.
Our method is implemented in an open-source, production-ready, Python package \texttt{mbank}\cite{mbank}, available on GitHub\footnote{
\href{https://github.com/stefanoschmidt1995/mbank}{stefanoschmidt1995/mbank}.}
and on the PyPI repository\footnote{
The package is distributed under the name \texttt{\href{https://pypi.org/project/gw-mbank/}{gw-mbank}}.
}.

The rest of this paper is devoted to the presentation and description of our methods and software package.
In Sec.~\ref{sec:methods} we present the details of our bank generation algorithm.
In Sec.~\ref{sec:validation} we assess the accuracy of our template placing method in all its parts.
Furthermore, we reproduce two banks available in the literature \cite{Harry:2017weg, Sakon:2022ibh} created with independent codes: this will be the topic of Sec.~\ref{sec:other_methods}.
To demonstrate the capabilities of \texttt{mbank}, in Sec.~\ref{sec:novel_applications}, we present two large banks covering ``exotic" regions of parameter space: a precessing bank and an IMBH bank with HM content. We also discuss some possible further applications of our normalizing flow model, including a study of the size of the precessing neutron star-black hole (NSBH) parameter space.
Finally, in Sec.~\ref{sec:improvements} we discuss some possible future development of our work and gather some final remarks in Sec.~\ref{sec:conclusion}.

Throughout the paper we will use the term ``standard" to refer to the searches for circularized, aligned-spin BBHs without imprints of HMs, currently conducted by the LIGO-Virgo-KAGRA collaboration.

\section{Methods} \label{sec:methods}

When searching for a BBH signal in GW data, it is customary to use a frequentist detection statistic~\cite{Creighton_book, Maggiore:2007ulw, Harry:2016ijz, Harry:2017weg}, which models the detector output to be composed of {\it gaussian} noise $n(t)$ and possibly a known GW signal $h(t)$.
Given some observed data $s(t)$, the detection statistic $\Lambda$ is a measure of the log probability ratio between the signal hypothesis $n+h$ and the noise hypothesis $n$:
\begin{equation}\label{eq:LL}
	\Lambda = \log\frac{p(s|n+h)}{p(s| n)}.
\end{equation}
For interferometric GW observatories such as LIGO and Virgo, the observed signal takes the following form:
\begin{equation}\label{eq:signal_model}
	h(t) = F_+(\delta, \alpha, \Psi) h_+(t;\theta) + F_\times(\delta, \alpha, \Psi) h_\times(t;\theta)
\end{equation}
The functions $F_+, F_\times$, also called antenna patterns, denote the interferometer response to the two polarizations of a GW. They depend on the sky location, parameterized by right ascension $\alpha$ and declination $\delta$, and on the polarization angle $\Psi$. 
For a BBH system, the two polarizations $h_+, h_\times$ depend on two BH masses ($m_1$, $m_2$), two 3-dimensional spins ($\mathbf{s}_1$, $\mathbf{s}_2$), the inclination angle $\iota$, the reference phase $\varphi$, the luminosity distance of the source $D_L$, the eccentricity $e$ of the orbit and the mean periastron anomaly $a$ \cite{Sathyaprakash_2009}.

Under the assumption of {\it Gaussian noise}, we can write down an explicit model for the likelihood and, after maximising over an overall amplitude factor, Eq.~\eqref{eq:LL} becomes \cite{Creighton_book, Maggiore:2007ulw, Harry:2016ijz}:
\begin{equation}\label{eq:LL_gauss}
	\Lambda = \frac{\left(\Re\scalar{s}{h}\right)^2}{\scalar{h}{h}} = \rescalar{s}{\hat{h}}^2
\end{equation}
where we introduced a {\it complex} scalar product between two vectors $a$, $b$:
\begin{equation} \label{eq:scalar_product}
	\scalar{a}{b} = 4 \int_{f_\text{min}}^{f_\text{max}} \!\!\!\! \d{f} \; \frac{\tilde{a}^*(f) \tilde{b}(f)}{S_n(f)}
\end{equation}
and the integral extends in a suitable frequency range $[f_\text{min}, f_\text{max}]$.
In this context, $S_n(f)$ is the frequency domain autocorrelation function of the noise, also called Power Spectral Density (PSD) and $\tilde{\phantom{a}}$ denotes the Fourier transform.
For ease of notation, we define ${\rescalar{a}{b} = \Re\scalar{a}{b}}$ and ${\hat{a} = \frac{a}{\rescalar{a}{a}}}$.

For any given observation time, a search aims to {\it maximize} the detection statistic $\Lambda$ with respect to all the parameters of the signal model. This maximized quantity is also called signal-to-noise ratio (SNR).
Depending on symmetry assumptions on the polarizations, one is able to maximize analytically over some (nuisance) parameters.
For the other quantities, a brute force approach is required, where the maximized $\Lambda$ is evaluated at each time on a large set of signal models, called a {\it template bank} \cite{PhysRevD.77.104017, Mukherjee:2018yra}.
Regardless of the nature of the signal, one is {\it always} able to maximise $\Lambda$ over sky-location (angles $\alpha$ and $\delta$), polarization angle $\Psi$ and luminosity distance $D_L$, which enters as an overall amplitude scaling.

The computation of the SNR as a function of time for a single template is known as {\it matched filtering} and has been implemented successfully as the first stage of several pipelines to search for GW signals \cite{Allen:2005fk, Privitera:2013xza, Usman:2015kfa, Capano:2016dsf, PhysRevD.95.042001, Nitz:2017svb, gstlal_paper2, Aubin:2020goo, Chu:2020pjv}. Modern pipelines can easily perform matched filtering on millions of templates and use the aggregated information to produce lists of GW candidates, ranked by their false alarm probability of occurrence in a noise only model.

For a circular non-precessing signal with no HM, it holds $\tilde{h}_+ \propto i\tilde{h}_\times$ and the maximization of Eq.~\eqref{eq:LL_gauss} over the nuisance parameters yields \cite{Maggiore:2007ulw}:
\begin{equation}\label{eq:std_snr}
	\max \Lambda = \lVert \scalar{s}{\hat{h}_+} \rVert^2 = \rescalar{s}{\hat{h}_+}^2 + \rescalar{s}{\hat{h}_\times}^2
\end{equation}
In this simple case, $\max\Lambda$ only depends on the two BH masses $m_1, m_2$ and the two z-components of the spins $s_{1z}, s_{2z}$ (4 quantities).

For the general case, where no particular symmetry is available, one obtains a different expression \cite{Capano:2013raa, Schmidt:2014iyl, Harry:2017weg}:
\begin{equation}\label{eq:symphony_snr}
	\max \Lambda = \frac{ \rescalar{s}{\hat{h}_+}^2 + \rescalar{s}{\hat{h}_\times}^2 -2\rescalar{\hat{h}_+}{\hat{h}_\times}\rescalar{s}{\hat{h}_\times}\rescalar{s}{\hat{h}_+}}{1- \rescalar{\hat{h}_+}{\hat{h}_\times}^2}
\end{equation}
In this case, $\max\Lambda$ depends on 12 parameters: they are the two BH masses $m_1, m_2$, the two three-dimensional spins $\mathbf{s}_1$, $\mathbf{s}_2$, the inclination angle $\iota$, the reference phase $\varphi$ and the eccentricity parameters $e, a$.
Unlike the ``standard" case, an analytical maximization does not remove the dependence of $\iota$ and $\varphi$, entering in $h_+, h_\times$.
Depending on the scope of a matched-filter search, a pipeline can use either Eq.~\eqref{eq:std_snr} or Eq.~\eqref{eq:symphony_snr} to filter the interferometer data with a template.

For the purpose of template placement, it is useful to think of the parameter space of BBH signals as a D-dimensional manifold $\mathcal{B}_D$, embedded in a large 12 dimensional manifold $\mathcal{B}$. Each point of the manifold corresponds to a GW signal. The number of dimensions $D$ depends on the BBH variables under consideration.
As the parameters that do not enter the interesting space can be freely neglected (i.e. set to $0$ or to a meaningful constant value), the manifold $\mathcal{B}_D$ is effectively a lower dimensional {\it projection} of the large manifold $\mathcal{B}$.

To place templates on $\mathcal{B}_D$, it is standard to equip the manifold with a distance (called {\it mismatch}), which also naturally defines a volume element at every point in space. The volume element defines the ``uniform" probability distribution according to the metric.
A random template bank will be populated by templates drawn from such distribution, until a certain coverage is reached. For this reason, our primary concern is to sample from the manifold and to check for coverage. To effectively do so, we rely on the three steps below:
\begin{enumerate}
	\item Construction of a metric approximation of the match between templates. This makes $\mathcal{B}_D$ a Riemannian manifold with a volume element
	\item Training of a normalizing flow model to sample from the manifold. 
	\item Placing the templates by sampling from the normalizing flow model and checking for coverage, following \cite{Coogan:2022qxs}.
\end{enumerate}
The rest of this section details the steps above.

\subsection{The metric} \label{sec:metric}

The definition of a metric on the manifold $\mathcal{B}_D$ provides a fast-to-compute approximation to the {\it mismatch} (distance) between templates and an estimation of the volume element at each point in the space.

Given two points of the manifold $\theta_1,\theta_2$, we define the overlap $\mathcal{O}(\theta_1,\theta_2, t)$ between normalized templates as:
\begin{widetext}
	\begin{align}\label{eq:overlap}
		\mathcal{O}(\theta_1,\theta_2, t) &= \frac{1}{1- \hat{h}_{+\times}(\theta_2)^2} 
		\biggl\{ \rescalarwide{\hat{h}_+(\theta_1)e^{i ft}}{\hat{h}_+(\theta_2)}^2 + \rescalarwide{\hat{h}_+(\theta_1)e^{i ft}}{\hat{h}_\times(\theta_2)}^2 \nonumber \\
		& -2h_{+\times}(\theta_2)\rescalarwide{\hat{h}_+(\theta_1)e^{i ft}}{\hat{h}_\times(\theta_2)}\rescalarwide{\hat{h}_+(\theta_1)e^{i ft}}{\hat{h}_+(\theta_2)} \biggl\}
	\end{align}
\end{widetext}
where $\hat{h}_+(\theta)e^{i ft}$ is the plus polarization $\hat{h}_+(\theta)$ translated by a constant time shift $t$ and $\hat{h}_{+\times}(\theta) = \rescalar{\hat{h}_+(\theta)}{\hat{h}_\times(\theta)}$.
The overlap amounts to the fraction of SNR recovered when filtering a signal $s=h_+(\theta_1)$ with a template evaluated at a point $\theta_2$ using Eq.~\eqref{eq:symphony_snr}.

In Eq.~\eqref{eq:overlap}, we choose to compare the $h_+$ polarization of the first template with both polarizations of the second template. We are forced to make such arbitrary choice by the fact that in general Eq.~\eqref{eq:symphony_snr} does depend on $F_+, F_\times$.
This creates an asymmetry between signal and template.
Thus, if we don't want the overlap to depend on two arbitrary combination coefficients, an arbitrary choice for the signal $s$ is needed.
Of course, any linear combination of $h_+(\theta_1)$ and $h_\times(\theta_1)$ works but we set $s = h_+(\theta_1)$ for computational convenience. Numerical studies show that replacing $h_+(\theta_1)$ with any linear combination does not have a large impact on the metric definition below.

In the case of a ``standard" search, $h_{+\times} = 0$ and ${\tilde{h}_\times = i \tilde{h}_+}$, hence the overlap simplifies to:
\begin{equation}\label{eq:overlap_NP}
\mathcal{O}(\theta_1,\theta_2, t) = \left|\scalar{\hat{h}_+(\theta_1)e^{i ft}}{\hat{h}_+(\theta_2)} \right|^2.
\end{equation}
Note that, since Eq.~\eqref{eq:std_snr} is symmetric\footnote{Indeed, for a ``standard" signal $s \propto h_+$, hence $\hat{s} = \hat{h}_+$, and Eq.~\eqref{eq:std_snr} does not depend on the antenna patterns functions, if $s$ is normalized.} between signal and template, the expression for the overlap in the ``standard" case is also symmetric. This means that an arbitrary choice on the signal composition is no longer needed, as was the case for Eq.~\eqref{eq:overlap}.

While all the literature available \cite{owen_metric, Messenger:2008ta, Prix:2007ks, Brown:2012qf, Roy:2017oul, Coogan:2022qxs, Hanna:2022zpk} relies on the expression in Eq.~\eqref{eq:overlap_NP} to derive the metric and addresses only the ``standard" case, we tackle the general case.

Closely following \cite{owen_metric}, can maximize the overlap Eq.~\eqref{eq:overlap} with respect to the time shift $t$ to obtain the {\it match} $\mathcal{M}(\theta_1,\theta_2)$ between templates evaluated at different points of the manifold:
\begin{equation}\label{eq:match}
	\mathcal{M}(\theta_1,\theta_2) = \max_t \mathcal{O}(\theta_1,\theta_2, t). \\
\end{equation}
The match has values in $[0,1]$ and trivially $\mathcal{M}(\theta_i,\theta_i) = 1$.

Even though in general the match is not symmetric and does not satisfy triangular inequality, we can use it to introduce a {\it distance} $d$ between two points on the D-manifold $\mathcal{B}_D$:
\begin{align}\label{eq:distance}
	d^2(\theta_1,\theta_2) \vcentcolon= 1 - \mathcal{M}(\theta_1,\theta_2).
\end{align}
The distance $d$ can then by approximated locally by a bilinear form $d_M$:
\begin{align}\label{eq:metric_definition}
	d_M^2(\theta_1,\theta_2) \vcentcolon= M_{ij}(\theta) \Delta\theta_i \Delta\theta_j \simeq 1 - \mathcal{M}(\theta_1,\theta_2).
\end{align}
The bilinear form $d_M$ is represented by a D-dimensional square matrix $M_{ij}(\theta)$, defined at each point of the manifold.

We identify $M_{ij}(\theta)$ to be the quadratic term of the Taylor expansion of ${d_M(\theta+\Delta\theta,\theta)}$ around $\Delta\theta\simeq 0$:
\begin{equation}\label{eq:metric_expression}
	M_{ij}(\theta) = - \frac{1}{2} \left( H_{ij} - \frac{H_{ti}H_{tj}}{H_{tt}} \right)
\end{equation}
where $H(\theta)$ is the Hessian of the overlap in Eq.~\eqref{eq:overlap}, a $D+1$ square matrix.
Note that the metric is positive definite (i.e. has positive eigenvalues).

A convenient expression for $H$ in terms of the gradients of the waveform is presented in App.~\ref{app:metric}, with the full expression given in Eqs.~\eqref{eq:H_tt_grad}-\eqref{eq:H_ij_grad}.
While identifying the metric with the Hessian is well motivated and yields reliable results, other definitions for $M_{ij}$ are possible; this is briefly discussed in App.~\ref{app:metric_definition}.

For most of the waveform models available, the gradients can be evaluated with finite difference methods. For a limited number of machine-learning based models \cite{Chua:2018woh, Khan:2020fso, Schmidt:2020yuu, Thomas:2022rmc, Tissino:2022thn}, the gradients are available analytically.

Equipped with the metric from Eq.~\eqref{eq:metric_expression}, the manifold $\mathcal{B}_D$ becomes a Riemannian manifold with line element:
\begin{equation}\label{eq:line_element}
	\d{s^2} = M_{ij}(\theta) \d{\theta_i} \d{\theta_j}.
\end{equation}
We can then use standard results from differential geometry to compute distances and volumes. In particular, the volume of a subset $\mathcal{T}$ of the manifold can be computed as:
\begin{equation}\label{eq:volume_tile}
	\text{Vol}(\mathcal{T}) = \int_\mathcal{T} \dvol{\theta}{D} \; \sqrt{\det M(\theta)}.
\end{equation}
where $\det M(\theta)$ is the determinant of the matrix $M_{ij}(\theta)$, also denoted as $|M|$.
Moreover, we introduce the uniform probability measure, such that $p(V) \propto \text{Vol}(V)$ for any $V\subseteq \mathcal{B}_D$. The measure has the following probability distribution function (PDF):
\begin{equation}\label{eq:pdf_uniform}
	p(\theta) \propto \sqrt{\det M(\theta)}.
\end{equation}
Samples from the uniform distribution tend to have a ``uniform" (i.e. constant) spacing, computed with the metric distance. Owing to this feature, the uniform distribution is a natural candidate to draw templates from.

\subsection{Sampling from the manifold} \label{sec:normalizing_flow}

To generate a random template bank, we need to sample points on the manifold $\mathcal{B}_D$ from Eq.~\eqref{eq:pdf_uniform}.
A simple way to do so is by means of a Markov Chain Monte Carlo (MCMC). However, this turns out to be unfeasibly expensive, since to obtain a single sample, the metric must be evaluated tens of times. For instance, to produce a bank with $\mathcal{O}(10^6)$ templates, $\mathcal{O}(10^7)$ metric evaluations are required.

To speed up the sampling, we introduce a {\it normalizing flow} model. As we will show below, in order to train the model $\mathcal{O}(10^5)$ metric evaluations are sufficient: this is a small fraction of the metric evaluations needed to run a MCMC.
Once trained, the normalizing flow model produces high quality samples from Eq.~\eqref{eq:pdf_uniform} in a small amount of time, effectively providing templates to populate a random template bank.

A normalizing flow model \cite{norm_flow, nflows_paper, Kobyzev_2021, Papamakarios_thesis} is a machine learning model widely used to reproduce and/or parameterize complicated probability distributions.
Mathematically, a flow is an {\it invertible} parametric function $\phi_W$ which is trained to map samples $\theta$ from an arbitrary probability distribution $p(\theta)$ to samples $\mathbf{x}$ from a multivariate standard normal distribution $\mathcal{N}(\mathbf{x}|0,\mathbf{1})$. The space of the $\mathbf{x}$ is sometimes referred to as {\it latent space}.
The parameters $W$ of the flow are set in such a way that:
\begin{equation}
	\mathbf{x} = \phi_W(\theta) \sim \mathcal{N}(\mathbf{x}|0,\mathbf{1}) \;\;\; \text{if} \;\;\;  \theta \sim p(\theta)
\end{equation}
In other words, a normalizing flow defines a parametric representation of a generic probability distribution $p(\theta)$, obtained by change of variables
\begin{equation}\label{eq:p_flow}
	p^\text{flow}_W(\theta) = \mathcal{N}(\phi_W(\theta)|0,\mathbf{1}) \; |\det J_{\phi_W}(\theta)|
\end{equation}
where $J_{\phi_W}$ is the Jacobian of the flow transformation $\phi_W$.
Sampling from $p^\text{flow}_W$ can then be easily done by sampling $\mathbf{x} \sim \mathcal{N}(\mathbf{x}|0,\mathbf{1})$ and obtaining $\theta$ from the inverse flow transformation: $\theta = \phi_W^{-1}(\mathbf{x})$.
Thus, given a target distribution, both the problems of sampling and of density estimation become tractable thanks to the normalizing flow model.

The flow transformation $\phi_W$ is built by {\it composing} $n_\text{layers}$ simple (invertible) transformations, each called a layer. Of course, depending on the application, a variety of options are available in the literature.
We build a layer by concatenating a linear transformation and a masked autoregressive layer \cite{MADE, MAF, MAF_bis} with $n_\text{hidden}$ hidden features.
A masked autoregressive layer implements the following transformation:
\begin{equation}
	T_{MADE}(\theta) = a(\theta)\theta+b(\theta)
\end{equation}
where the coefficients $a(\theta), b(\theta)$ are computed by (masked) autoencoders with $n_\text{hidden}$ hidden features.

In our case, the target probability distribution has support in the rectangle $[\theta_\text{min}, \theta_\text{max}]$, while the base distribution of the flow (a Gaussian) has support in $\mathbb{R}^D$. We implement the change of support explicitly by introducing the following transformation $T_0(\theta): [\theta_\text{min}, \theta_\text{max}] \to \mathbb{R}^D$ as the first layer of the flow:
\begin{equation}\label{eq:first_transform}
	T_0(\theta) = 0.5 \log \frac{1 + y}{1 - y} \;\;\; \text{with} \;\;\; y = \frac{2\theta - \theta_\text{min} - \theta_\text{max}}{\theta_\text{max}- \theta_\text{min}}
\end{equation}
where the fraction above is intended as element-wise division.\footnote{
Note that the inverse $T_0^{-1}$ of the transformation takes a simple form: $\frac{1}{2} [\text{tanh}(T_0(\theta))(\theta_\text{max} - \theta_\text{min})+\theta_\text{max}+ \theta_\text{min}]$, where again the multiplication is intended as element-wise.
}
This transformation maps the rectangle $[\theta_\text{min}, \theta_\text{max}]$ into the plane. Then the remaining transformations only need to implement a change in probability density and not in the support of the distribution, making the loss function optimization easier.

The flow probability distribution $p^\text{flow}_W(\theta)$ is trained to closely reproduce a given probability distribution $p^\text{target}(\theta)$.
During the training, the weights $W$ of the flow are set by minimizing a loss function $\mathcal{L}_\phi(W)$, which measures the discrepancy between $p^\text{target}$ and $p^\text{flow}_W$. The minimization is performed by gradient descent.
In our case, $p^\text{target} \propto \sqrt{\det M}$, with an unset normalization.

Depending on the nature of the data, several loss functions are available.
If {\it samples} from the target distribution are available, the loss function is defined as the forward Kullback–Leibler (KL) divergence between the target distribution $p^\text{target}(\theta)$ and the one defined by the flow in Eq.~\eqref{eq:p_flow}:
\begin{align}
	\mathcal{L}^{KL}_\phi(W) = - \mathbb{E}_{p^\text{target}(\theta)} [\log p^\text{flow}_W] + \text{const.}
\end{align}
where the expected value is computed using empirical samples from $p^\text{target}(\theta)$ to provide a Monte-Carlo estimation of the loss function.

In our situation however, we do not have access to such samples (indeed, we are training the flow precisely to avoid sampling!) but we are only able to evaluate $p^\text{target}$ up to a constant scaling factor.
For this reason, we treat the training as a {\it regression} problem, rather than a density estimation problem, and we use the following loss function:
\begin{align}\label{eq:loss_mse}
	\mathcal{L}_\phi(W) &= \frac{1}{N} \sum_{i=1}^N \left(\log p^\text{flow}_W(\theta_i) - \log p^\text{target}(\theta_i) \right)^2 \nonumber\\
						&= \frac{1}{N} \sum_{i=1}^N \left(\log p^\text{flow}_W(\theta_i) - \log\sqrt{|M(\theta_i)|} + C \right)^2
\end{align}
where the sum runs on a dataset of $N$ points:
\begin{equation}
	\{(\theta_i, \sqrt{|M(\theta_i)|})\}_{i=1}^N
\end{equation}
Our experiments show that ${N \simeq 5 \times 10^5}$ is adequate in most cases.

In Eq.~\eqref{eq:loss_mse}, $C$ is a {\it trainable} constant, which sets the normalization of $p^\text{target} = e^{-C} \sqrt{|M|}$ on the domain of interest.
Although not strictly needed, it can have a large impact on the flow performance, since it constrains the values of $\sqrt{|M(\theta)|}$ to a scale which is easier to learn by the normalizing flow.
Some heuristics suggest initializing the constant to the $90^\text{th}$ percentile of the values $\log \sqrt{|M(\theta)|}$ stored in the dataset.
As shown in App.~\ref{app:parameter_space_volume}, the constant can be used to compute (an approximation to) the volume of the parameter space $\mathcal{V}$ in Eq.~\eqref{eq:volume_tile}.

The values of $\theta_i$ in Eq.~\eqref{eq:loss_mse} are obtained by sampling the masses $m_1, m_2$ from
\begin{equation}\label{eq:sampling_dataset}
	p(\mathcal{M}_c, \eta) \propto \mathcal{M}_c^{10/3} \eta^{8/5}
\end{equation}
where $\mathcal{M}_c = \frac{(m_1m_2)^{3/5}}{(m_1+m_2)^{1/5}}$ is the chirp mass and $\eta = \frac{m_1m_2}{(m_1+m_2)^2}$ is the symmetric mass ratio.
All other quantities are sampled from a uniform distribution in the coordinates.

Eq.~\eqref{eq:sampling_dataset} defines a flat distribution on the chirptime parameters $\tau_0$ and $\tau_3$ \cite{Cokelaer:2007kx}. Indeed, it can be shown that for a nonspinning binaries, the metric expressed in the chirptime coordinates is approximately flat \cite{Sathyaprakash:1991mt, Dhurandhar:1992mw}, and that Eq.~\eqref{eq:sampling_dataset} represents a first order approximation to the true metric.
Sampling from Eq.~\eqref{eq:sampling_dataset}, ensures a high quality training set, where the distribution of the training points is reasonably close to the target distribution\footnote{Indeed more samples are present at low chirp mass, which is where the metric determinant tends to have larger values due to longer waveforms (for a constant starting frequency). Hence, a consistent bias in the low mass region is largely penalized in the loss function due to more samples in the dataset at low mass.}.

During the training we halve the learning rate each time the validation loss does not improve more than a given threshold after a given number of iterations. This procedure finds local minima better in the loss function. We also apply early stopping, to avoid useless gradient descent iterations.

The training of the normalizing flow usually takes $\mathcal{O}(30 \text{ minutes})$. On the other hand, from one to a few hours are needed to generate a dataset of $\mathcal{O}(10^5)$ points, depending on the dimensionality of the manifold and on the waveform approximant.
This is the bulk of the cost of generating a template bank: the random template placing takes only a few minutes.

\subsection{Random template placing} \label{sec:template_placing}

As customary, the input parameter controlling the average spacing and number of templates is the {\it minimal match} $MM$. It is defined as the minimum tolerable match that a random signal (inside the relevant parameter space) must have with its nearest templates in the bank.
Of course, during the template placement, we only consider the match between templates on the same manifold, while the quantity can be used also to compare waveforms on different manifolds.

To generate our random template bank, following \cite{Messenger:2008ta}, we add random templates to the bank until a satisfactory coverage is achieved. The coverage is checked using a procedure that closely matches \cite{Coogan:2022qxs}.
The templates are sampled from the normalizing flow in Eq.~\eqref{eq:p_flow}, which, as discussed above, is trained to target Eq.~\eqref{eq:pdf_uniform}. This choice makes sure that the templates are spread as ``uniformly as possible" across the manifold.

One point of the space $\theta$ is said to be {\it covered} by the bank if there is at least one template $\theta_T$ in the bank, whose squared metric distance (mismatch) as given in Eq.~\eqref{eq:metric_definition} is at most $1 - MM$ or:
\begin{equation}
	d^2_M(\theta, \theta_T)<1 - MM.
\end{equation}
The covering fraction $\hat{\eta}$ of a given region $\mathcal{T}$ of the parameter space is then defined as the fraction of volume covered by the bank:
\begin{equation}\label{eq:coverage}
	\hat{\eta}(\mathcal{T}) = \frac{1}{\text{Vol}(\mathcal{T})} \int_\mathcal{T} \dvol{\theta}{D} \; \sqrt{\det M(\theta_i)} \; c(\theta).
\end{equation}
where $c(\theta)$ is an indicator function:
\begin{equation}
	c(\theta) = \left\{
                \begin{array}{ll}
                  1 \;\; \text{if $\theta$ is covered by the bank}\\
                  0 \;\; \text{otherwise}
                \end{array}.
              \right.
\end{equation}
We do not require that the space is fully covered but we only require that it is covered with probability $\eta$. This means that we terminate the bank construction when the covering fraction $\hat{\eta} \geq \eta$.

To provide a sensible estimate of the covering fraction $\hat{\eta}$, we perform a Monte Carlo estimation of the integral in Eq.~\eqref{eq:coverage} \cite{Coogan:2022qxs}:
\begin{equation}\label{eq:coverage_estimate}
	\hat{\eta}(\mathcal{T}) \simeq \frac{1}{N_\text{livepoints}} \sum_i c(\theta_i)
\end{equation}
where the $N_\text{livepoints}$ samples $\theta_i \sim p^\text{flow}$ are sampled from the normalizing flow and are called {\it livepoints}.
Note that in Eq.~\eqref{eq:coverage_estimate}, we don't compute volumes using the volume element $\sqrt{\det M}$ itself but rather its normalizing flow approximation.

In practice, while the templates are being added to the bank, the distance between each livepoint is computed. If the i-th livepoint is close enough to the newly added template, it will be removed from the set of livepoints and a running estimate of $\hat{\eta}(\mathcal{T})$ will be updated.
The estimation of the covering fraction $\hat{\eta}$ has standard deviation \cite[App. A]{Coogan:2022qxs}:
\begin{equation}\label{eq:variance_coverage}
	\sigma_{\hat{\eta}} = \sqrt{\frac{\eta(1-\eta)}{N_\text{livepoints}-1}}
\end{equation}
which suggests using a large number of livepoints for better estimation.
In \cite{Coogan:2022qxs}, the authors typically choose $\eta = 0.9$ and $N_\text{livepoints} = 2000$.

Since the method does not check for distances between templates, it can overcover the space (as also reported in \cite{Messenger:2008ta, Coogan:2022qxs}), especially for a low number of dimensions.
Despite this, it is very fast and provides a reliable bank at a cheap computational and memory cost.
Moreover, as argued in \cite{Messenger:2008ta, Allen:2021yuy, Allen:2022lqr}, for a large number of dimensions, the banks generated by the random method provide close to optimal performance.

As a final remark, we note that for the purpose of computing the covering fraction, the templates do not need to be stored, which enables the algorithm to run with a very low memory footprint.
As exemplified in Sec.~\ref{sec:other_applications}, this allows to study the number of templates required to cover a particular region of the parameter space, providing invaluable pieces of information useful to plan a GW search.

\begin{figure}[t]
	\centering
	\includegraphics[scale = 1.]{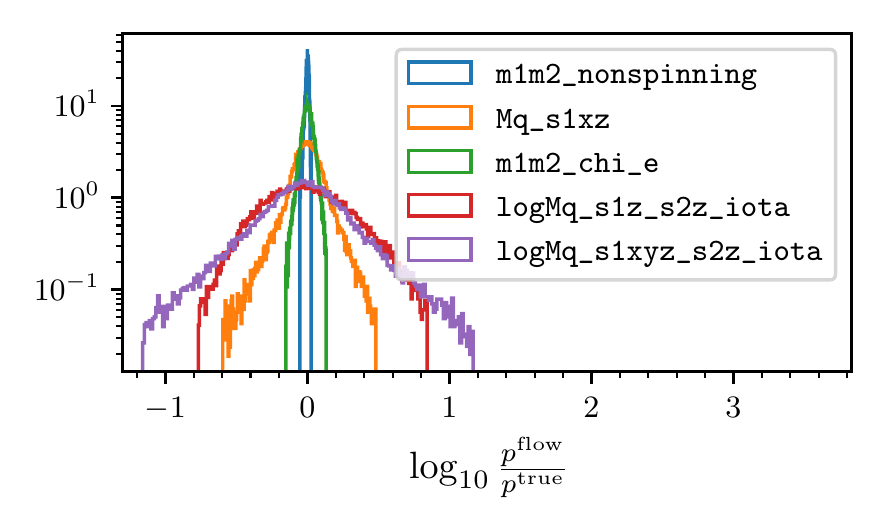}
	\caption{Study of the accuracy for several normalizing flow, trained on different manifolds. For each manifold, we compute the logarithmic ration $\log_{10}\frac{p^\mathrm{flow}}{p^\mathrm{true}}$ between the PDF computed by the flow and the true one. We use $40000$ test points from the validation set of each manifold. Details on the manifold considered are reported in Tab.~\ref{tab:flow_validation}.}
	\label{fig:flow_validation}
\end{figure}

\begin{table}
	\begin{tabular}{c c c c} 
	 \phantom{manifold} & Parameter space & $D$ & Architecture \\ 
	 \toprule
	 \texttt{m1m2\_nonspinning} &
	 	\begin{tabular}{@{}l@{}} $m_1,m_2\in [1, 200] \mathrm{M_\odot}$ \\ $q\in [1,30]$ \\ $f \in [15, 1024] \SI{}{Hz}$ \\
	 	\texttt{IMRPhenomD} \cite{Khan:2015jqa} \end{tabular} & 2 & 60 60 30\\
 	\addlinespace[3pt]
 	\cdashline{1-4}
 	\addlinespace[3pt]
	 \texttt{Mq\_s1xz} & 
	 	\begin{tabular}{@{}l@{}} $M\in [25, 100] \mathrm{M_\odot}$ \\ $q\in [1,5]$  \\ $s_1\in [0, 0.99]$  \\ $\theta_1 \in [0, \pi]$  \\ $f \in [15, 1024] \SI{}{Hz}$\\
	 	\texttt{IMRPhenomXP} \cite{Pratten:2020ceb} \\ \end{tabular} & 4 & 70 70\\
 	\addlinespace[3pt]
	\cdashline{1-4}
	\addlinespace[3pt]
	 \texttt{m1m2\_chi\_e} & 
		\begin{tabular}{@{}l@{}} $m_1,m_2\in [1, 50] \mathrm{M_\odot}$ \\ $q\in [1,20]$  \\ $\chi_\text{eff} \in [-0.99, 0.99]$ \\ $e\in [0, 0.5]$ \\  $f \in [10, 1024] \SI{}{Hz}$ \\ \texttt{EccentricFD} \cite{lalsuite} \\ \end{tabular} & 4 & 60 60 60\\
	\addlinespace[3pt]
	\cdashline{1-4}
	\addlinespace[3pt]
		\begin{tabular}{@{}c@{}}
			\texttt{logMq\_s1z\_s2z\_iota} \\ (with HM) \\
		\end{tabular} & 
	 	\begin{tabular}{@{}l@{}} $m_1,m_2\in [50, 300] \mathrm{M_\odot}$ \\ $M\in [100, 400] \mathrm{M_\odot}$ \\ $q\in [1,10]$  \\ $s_{1z},s_{2z}\in [-0.99, 0.99]$  \\ $\iota \in [0, \pi]$  \\ $f \in [10, 1024] \SI{}{Hz}$\\
	 	\texttt{IMRPhenomXP} \cite{Pratten:2020ceb} \\ \end{tabular} & 5 & 20 60 60\\
 	\addlinespace[3pt]
	\cdashline{1-4}
	\addlinespace[3pt]
		\texttt{logMq\_s1xyz\_s2z\_iota} &
	 	\begin{tabular}{@{}l@{}} $m_1,m_2\in [1, 100] \mathrm{M_\odot}$ \\ $M\in [2, 150] \mathrm{M_\odot}$ \\ $q\in [1,20]$  \\ $s_1\in [0, 0.99]$  \\ $\theta_1 \in [-\pi, \pi]$ \\ $\phi_1 \in [0, \pi]$ \\ $s_{2z}\in [-0.99, 0.99]$  \\ $\iota \in [0, \pi]$  \\ $f \in [15, 1024] \SI{}{Hz}$\\ \texttt{IMRPhenomXHM} \cite{Garcia-Quiros:2020qpx} \\ \end{tabular} & 7 & 100 60 60 60\\
	 \bottomrule
	\end{tabular}
	\caption{Details of the manifold considered for the validation of the normalizing flow model in Fig.~\ref{fig:flow_validation}. For each manifold, we report the variables being sampled together with their ranges. We also list the the frequency range considered, the waveform approximant used, the number of dimensions $D$ of the manifold as well as the number of hidden features for each layer of the flow. }
	\label{tab:flow_validation}
\end{table}

\begin{figure}[t]
	\centering
	\includegraphics[scale = 1.]{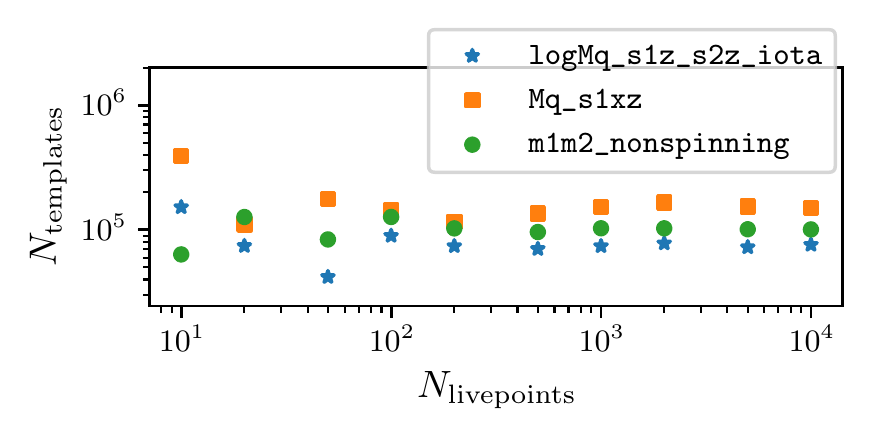}
	\caption{Validation of the random template placement algorithm. For three of the manifolds introduced in Tab.~\ref{tab:flow_validation}, we plot the number the number of templates $N_\text{templates}$ of a random template bank as a function of the number of livepoints $N_\text{livepoints}$ used to estimate the covering fraction. For each template bank, we set $\eta = 0.9$ and $MM = 0.97$. }
	\label{fig:template_placement_validation}
\end{figure}

\section{Validation} \label{sec:validation}

In this section, we assess the performance of the two key ingredients of our template bank generation algorithm, namely the normalizing flow model and the random placement algorithm.
Our goal is to understand the limitations of our algorithm as well as to make an informed choice of the various hyperparameters that impact the quality of the template bank.

We will consider different manifolds, which will be named with a string that lists the manifold coordinates. The coordinates are grouped by mass coordinates, spin coordinates, (eventual) eccentricity coordinates (i.e. $e$ and $a$) and (eventual) angles coordinates (i.e. $\iota$ and $\varphi$).
Consequently, a string has the format \texttt{Masses\_Spin1\_Spin2\_Eccentricity\_Angles}.

Valid options for the mass coordinates are \texttt{m1m2} which uses $m_1$ and $m_2$ as coordinates, \texttt{Mq} which uses total mass $M = m_1+m_2$ and mass ratio $q = m_1/m_2 >1$, and \texttt{logMq} which uses $\log_{10}M$ instead of $M$.
Similarly, other variables are listed by their names.
The manifold with spin label \texttt{chi} uses the effective spin parameter
\begin{equation}\label{eq:chieff}
	\chi_\text{eff} = \frac{m_1 s_\text{1z} + m_2 s_\text{2z}}{m_1 + m_2}
\end{equation}
as coordinate. Since $\chi_\text{eff}$ is degenerate in the two spins, we choose to set $s_\text{1z} = s_\text{2z} = \chi_\text{eff}$ and all the other spin components to $0$.

If more than one spin coordinate is given for a given BH, the spin vector $\mathbf{s}$ will be parameterized in spherical coordinates with magnitude ${s \in [0,1)}$ and angles ${\theta \in [-\pi,\pi]}$ and ${\varphi \in [0, \pi]}$ as follows:
\begin{align}
	s_\text{x} & = s \sin\theta \cos\phi \\
	s_\text{y} & = s \sin\theta \sin\phi \\
	s_\text{z} & = s \cos\theta.
\end{align}
Note that the angle $\theta$ controls the amount of precession. With $\theta = 0, \pm \pi$ the spin has only a z component (i.e., is aligned with the orbital angular momentum), while for $\theta = \pm\pi/2$ there is maximal precession, as the spin vector only has an in-plane component.

\subsection{Normalizing flow validation} \label{sec:flow_validation}

To study the accuracy of the normalizing flow model in reproducing $\sqrt{|M|}$, we consider five manifolds. The manifolds are listed in Tab.~\ref{tab:flow_validation}, together with the region of the parameter space they cover. We also report the waveform approximant used as well as the frequency range where the metric is computed.
The manifolds were chosen to have a variety of number of dimensions $D$ and to cover a broad ranges of physical scenarios (nonspinning, aligned-spins, precession, HM, and eccentric orbits).

For each manifold we generate a dataset of $3\times 10^5$ points and we compute the (log) value of the PDF in Eq.~\eqref{eq:pdf_uniform}. We then train a normalizing flow model on each of the datasets.
The architecture of each flow is also reported in Tab.~\ref{tab:flow_validation}.

Fig.~\ref{fig:flow_validation} shows a histogram with the accuracy of the normalizing flow reconstruction of the PDF on each manifold. This is quantified by $\log_{10}\frac{p^\mathrm{flow}}{p^\mathrm{true}}$, which measures the logarithmic ratio between the two PDFs.

Overall, the accuracy of the flow is (almost) always contained within one order of magnitude. Whether a similar error is acceptable for the purpose of template placement needs to be checked on a case-by-case basis with an injection study, as discussed in Sec.~\ref{sec:other_methods}.

We note that all histograms are well-centered around $0$, showing that the flow does not have a systematic bias. Moreover, the accuracy tends to be higher for low-dimensional manifolds. Indeed, low dimensional manifolds present an easier learning task for the flow.

The manifold \texttt{logMq\_s1xyz\_s2z\_iota} shows the largest spread in accuracy, as it is the largest dimensional manifold being considered. Note that it parameterizes a huge parameter space, which cannot be realistically covered by a template bank. Hence, as a realistic bank will necessarily cover a subset of the manifold, a flow trained on that smaller parameter space will most certainly show better accuracy, due to an easier regression task.

Finally, we see that the flow trained on the eccentric manifold \texttt{m1m2\_chi\_e} has remarkably good performance. This can be explained by the fact that the approximant \texttt{EccentricFD} \cite{lalsuite} used is analytical. This ensures very smooth behaviour across the parameter space, which can be easier for the normalizing flow model to learn.

\subsection{Template placement performance} \label{sec:template_placement}

As already stated, the template placement method in use closely matches the one introduced in \cite{Coogan:2022qxs}.
The main novelty introduced here is sampling with the normalizing flow as opposed to rejection sampling.

For the random placement method, there are two parameters to tune that affect the final bank size. They are the number of livepoints $N_\text{livepoints}$ and the covering fraction $\eta$.
The authors of \cite{Coogan:2022qxs} make an extensive investigation on how the bank size depends on such quantities and we do not repeat such in-depth studies here.

We limit ourselves to examining the convergence of the template number $N_\text{templates}$ as a function of $N_\text{livepoints}$ (see \cite[Fig.~4 (right)]{Coogan:2022qxs}) in the case of manifolds with precessing and HM signals.
For the study, we chose the manifolds \texttt{m1m2\_nonspinning}, \texttt{Mq\_s1xz} and \texttt{logMq\_s1z\_s2z\_iota} introduced in Sec.~\ref{sec:flow_validation} (see also Tab.~\ref{tab:flow_validation}). The second manifold covers a precessing parameter space, while the metric on the latter manifold is computed with an HM approximant \cite{Garcia-Quiros:2020qpx}.

We present our results in Fig.~\ref{fig:template_placement_validation}, where the number of templates is computed with a covering fraction $\eta = 0.9$ with varying $N_\text{livepoints}$.
In all cases the number of templates converges to a constant value as $N_\text{livepoints}$ increases. Already $\sim 500$ livepoints are enough to provide an accurate estimation of the bank size.
Our results are consistent with the findings of \cite{Coogan:2022qxs}, which we further extend to higher-dimensional manifolds.

\section{Comparison with other bank generation methods} \label{sec:other_methods}

We compare the output of \texttt{mbank} with two banks available in the literature, generated with two different methods.
The first bank is a non-spinning HM bank \cite{Harry:2017weg}, covering the high mass region of the BBH parameter space. The bank was generated using the stochastic placement algorithm, as implemented in the code \texttt{sbank} \cite{Ajith:2012mn}.
The second bank is the aligned-spin bank \cite{Sakon:2022ibh} currently in use by the \texttt{GstLAL} pipeline \cite{PhysRevD.95.042001, gstlal_paper2} for the fourth observing run (O4) of the LIGO-Virgo-Kagra collaboration. It was generated using the \texttt{manifold} \cite{Hanna:2022zpk} metric template placement algorithm called and covers a very wide mass range in the BNS and BBH parameter space.
Both banks have a minimal match $MM$ requirement of $0.97$.

In much of what follows we will measure the coverage of a bank. To do so, we randomly extract a number of simulated signals and, for each of them, we compute the maximum match with the templates of the bank. The latter quantity is called {\it fitting factor} $FF$ which, for a simulated signal characterized by orbital parameters $\theta$, it is defined as:
\begin{equation}\label{eq:FF}
	FF(\theta) = \max_{\theta^\prime \in \text{bank}} \mathcal{M}(\theta, \theta^\prime)
\end{equation}
Clearly, the match is computed using Eq.~\eqref{eq:symphony_snr}.

Borrowing the jargon of GW searches, we call {\it injections} the simulations for which we evaluate the fitting factor. In a real search, such signals would be added to the interferometer's data (i.e. injected) to measure the performance of the pipeline: the fitting factor measures the fraction of SNR lost due to the discreteness of the template bank.


\begin{table}
	\begin{tabular}{c c c c} 
	 \phantom{Name} & Parameter space & \multicolumn{2}{c}{
		\begin{tabular}{c c} \multicolumn{2}{c}{Size}  \\ Original & \texttt{mbank} \\ \end{tabular}	 
	 } \\ 
	 \toprule
	 HM bank \cite{Harry:2017weg} & \begin{tabular}{@{}l@{}} $M\in [50, 400] \mathrm{M_\odot}$ \\ $q\in [1,10]$  \\ $\iota\in [0,\pi]$ \\ $\varphi\in [0,2\pi]$   \\ \texttt{IMRPhenomXHM} \cite{Garcia-Quiros:2020qpx} \\ \end{tabular} & 20500 & 58932 \\
 	\addlinespace[3pt]
	\cdashline{1-4}
	\addlinespace[3pt]
	 ``All-sky" bank \cite{Sakon:2022ibh} & \begin{tabular}{@{}l@{}} $m_1,m_2\in [1, 200] \mathrm{M_\odot}$ \\ $q\in [1,20]$  \\ $\chi_\text{eff}\in [-0.99,0.99]$   \\ \texttt{IMRPhenomD} \cite{Khan:2015jqa}\\ \end{tabular} & $1.8 \times 10^6$  & $1.3 \times 10^6$  \\
	 \bottomrule
	\end{tabular}
	\caption{Details of the two banks available in the literature that we reproduce with our code. For each bank, we indicate the parameter space considered and the approximant used. We also compare the number of templates of the banks obtained with the different methods. }
	\label{tab:bank_comparison}
\end{table}

\begin{figure}[t]
	\centering
	\includegraphics[scale = 1.]{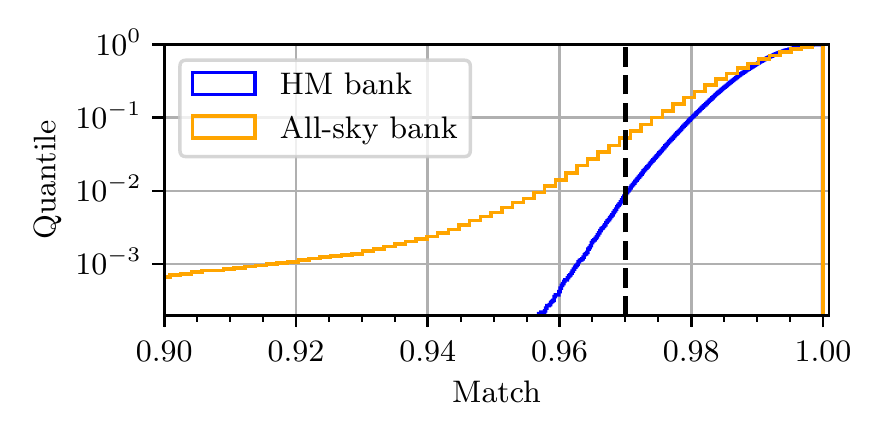}
	\caption{Fitting factor studies for the two template banks introduced in Sec.~\ref{sec:other_methods}. As discussed in Sec.~\ref{sec:HM_comparison} and~\ref{sec:all_sky_comparison} respectively, ``HM bank" is designed to reproduce \cite{Harry:2017weg} and targets high mass non-spinning systems with HM content, while the ``All-sky bank" bank covers aligned-spin systems (without HM) over a broad mass range, following \cite{Sakon:2022ibh}. We report the cumulative histogram of the fitting factors of $10^5$ injections samples across the parameter space.}
	\label{fig:test_banks_hist}
\end{figure}

\begin{figure}
	\includegraphics[scale = 1.]{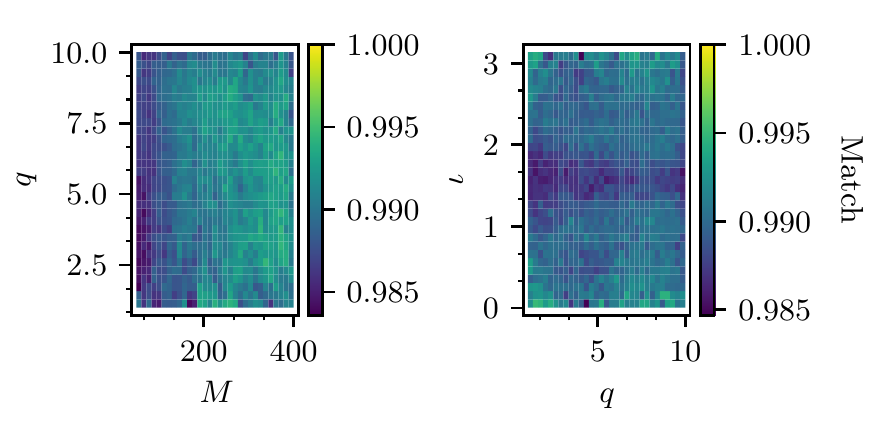}
	\caption{Validation of the ``HM bank", generated with our code and designed to reproduced \cite{Harry:2017weg}. For each two dimensional bin, we report the median fitting factor of $10^5$ injections covering the parameter space, as described in the text.}
	\label{fig:symphony_HM_injections}
\end{figure}

\begin{figure}
	\includegraphics[scale = 1.]{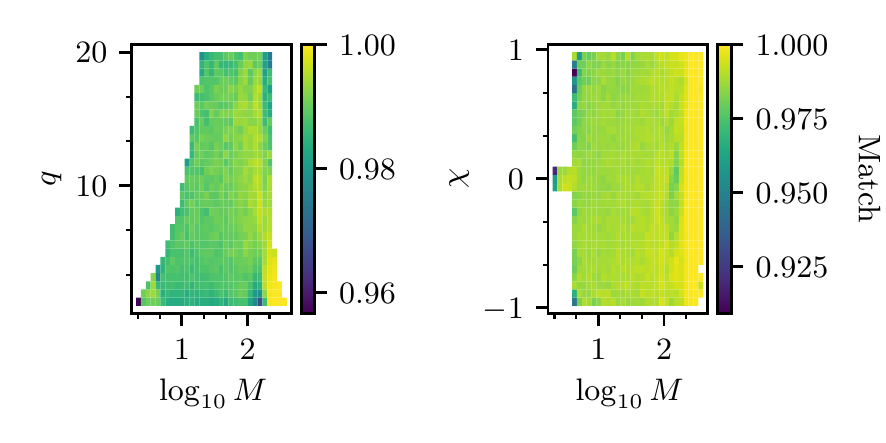}
	\caption{Validation of the ``All-sky bank", generated with our code and designed to reproduced \cite{Sakon:2022ibh}. For each two dimensional bin, we report the median fitting factor of $10^5$ injections covering the parameter space, as described in the text.}
	\label{fig:bank_O4_injections}
\end{figure}

\subsection{A non-spinning HM template bank} \label{sec:HM_comparison}

The non-spinning HM bank described in \cite{Harry:2017weg} covers systems with total mass $M$ in the range $[50, 400] \mathrm{M_\odot}$ and mass ratio $q\in [1,10]$. It also includes the inclination angle $\iota$ and reference phase $\varphi$ of the system, both covering the whole possible spectrum of values $\iota\in [0,\pi]$ and $\varphi\in [0,2\pi]$.
The authors use the analytical ``zero-detuning high power" PSD \cite{OLD_PSDs} and consider a low frequency cutoff $f_\text{min} = \SI{10}{Hz}$.

As already noted, they use the state-of-the-art code \texttt{sbank} \cite{Ajith:2012mn, PhysRevD.80.104014}.
The method is very accurate and known to provide effective coverage with a low number of templates. Of course, this comes at a large up-front computational cost to construct the bank.

To reproduce this bank, we place templates on the manifold \texttt{logMq\_nonspinning\_iotaphi}, with coordinates $\log_{10}M$, $q$, $\iota$ and $\varphi$. We use the same PSD and coordinate ranges as the original bank. We refer to our bank as ``HM bank".
We train a normalizing flow model with $4$ layers with $60, 60, 60, 10$ hidden features respectively and we choose $N_\text{livepoints} = 2000$ and a covering fraction $\eta = 0.8$.
Our bank has $58932$ templates and took a few hours to generate; the original bank is reported to have $20500$ templates.
All information is summarized in Tab.~\ref{tab:bank_comparison}.
We perform an injection study, drawing $10^5$ signals uniformly sampled in $\log M, q, \cos\iota$ and $\varphi$. The results of such study are reported in Fig~\ref{fig:test_banks_hist} and Fig~\ref{fig:symphony_HM_injections}.

First we note that our bank successfully covers the parameter space, with only $1\%$ of injections found with fitting factor below $0.97$ and less than $1\%$ with fitting factor below $0.96$. The coverage of the bank is similar to that of \cite{Harry:2017weg}.
In Fig.~\ref{fig:symphony_HM_injections}, we observe that the coverage is uniform across the space, i.e. we do not see regions where the fitting factor is significantly different from the others.

Comparing the number of templates, it is striking that our bank has almost three times more templates than the original template bank.
As no template rejection is done during the random bank construction, there is no control over templates being too close to each other. For this reason, an over-coverage of the space is inherent to the random template placement and is also reported in \cite{Messenger:2008ta,Coogan:2022qxs}. This problem can be addressed in future work, as discussed in Sec.~\ref{sec:improvements}.

\subsection{An ``All-sky" template bank} \label{sec:all_sky_comparison}

The aligned-spin bank (with no HMs) introduced in \cite{Sakon:2022ibh} covers a broad mass range, with systems with component masses $m_1, m_2 \in [1,200]\,\mathrm{M_\odot}$. The spins of the two objects are constrained to be equal to each other\footnote{This choice reduces the dimension of the manifold, without compromising the template bank accuracy.}, ${s_\text{1z} = s_\text{2z} = \chi_\text{eff}}$, spanning the range $[-0.99, 0.99]$.
The authors set an upper limit to the mass ratio $q<20$. Moreover, for objects with component mass $m<3\,\mathrm{M_\odot}$, they limit $\chi_\text{eff}$ in the range $[-0.05, 0.05]$\footnote{This is motivated by astrophysical considerations. Objects with masses smaller than $3\,\mathrm{M_\odot}$ are likely to be neutron stars and such objects are believed to develop only mild rotations~\cite{Zhu:2017znf}.}.
The authors use the Advanced LIGO O4 Design PSD (with $\SI{190}{Mpc}$ range) \cite{O4_PSDs} and consider a low frequency cutoff ${f_\text{min} = \SI{10}{Hz}}$.

The comparison with \cite{Sakon:2022ibh} is particularly interesting, since the bank is also produced with a metric template placement, implemented in the \texttt{manifold} code \cite{Hanna:2022zpk}. \texttt{manifold} uses a geometric approach, where the parameter space is iteratively split into (hyper)rectangles along the coordinates, until the volume of each rectangle reaches a sufficiently small value that it can be covered by a single template.

As summarized in Tab.~\ref{tab:bank_comparison}, we construct a bank to cover the parameter space used in~\cite{Sakon:2022ibh} over the manifold \texttt{m1m2\_chi}, sampling the coordinates $m_1$, $m_2$ and $\chi_\text{eff}$. We may refer to our bank as the ``All-sky" bank.
To produce our ``All-sky" bank, we trained three different normalizing flows in different regions of the parameter space. A first normalizing flow covers the BBH region with $m_1 \in [3,200]\,\mathrm{M_\odot}$, with $\chi_\text{eff} \in [-0.99, 0.99]$. A second one covers the BNS region, covering the manifold, ${(m_1, m_2, \chi_\text{eff}) \in [1,3]\,\mathrm{M_\odot}\times[1,3]\,\mathrm{M_\odot}\times [-0.05, 0.05]}$.
A third normalizing flow specializes in the high mass region, characterized by $m_1, m_2 \in [100,200]\,\mathrm{M_\odot}$.
Indeed, at high masses, the template density is so low that hardly any livepoint is sampled, which results in dramatic undercoverage. An appropriate coverage is enforced by the third normalizing flow, which places $\mathcal{O}(3000)$ templates in the region as opposed to {\it zero} templates placed by the first flow.
The additional coverage at high masses is manifest in Fig.~\ref{fig:bank_O4_injections}, as discontinuity in the fitting factor for ${m_1,m_2>100\,\mathrm{M_\odot}}$.

All the three normalizing flow models are made of $5$ layers of $10$ hidden features each.
Three templates banks are generated using each normalizing flow and they are merged together afterwards.
For the template placement we set $N_\text{livepoints} = 2000$ and covering fraction $\eta = 0.95$.
The resulting bank has $1326805$ templates.

The bank generation took around three hours, with most of the computing time spent on the dataset generation (i.e. on expensive metric computation). If needed, the dataset generation can be easily parallelized using \texttt{mbank}, hence reducing significantly the bank generation time.
Relying on parallel execution, \cite{Sakon:2022ibh} reported a generation time of minutes.

To validate our bank, we generate an injection set with $10^5$ injections, with the logarithm of the masses uniformly sampled.
Results of our injections studies are reported in Fig~\ref{fig:test_banks_hist} and Fig~\ref{fig:bank_O4_injections}.
Note that our injection set is different from the ones used in \cite{Sakon:2022ibh}.

In Fig~\ref{fig:test_banks_hist}, we see that $\sim 5\%$ of the injections have a match below $0.97$. The low fitting factor injections are mostly located around the low mass corners of the bank, clustered on the low mass end of the BNS region and in the high spin - low mass edge of the BBH region.
Inside the template bank and on the high mass end of the parameter space, satisfactory coverage is achieved.
Our results suggest that \texttt{mbank} struggles to accurately cover the ``narrow" corners of the parameter space. Nevertheless, this is a common problem that has been observed with other placement methods as well, and several strategies have been proposed to cope with it. Within our framework, the simplest option would be to extend the boundaries of the bank at low masses, thus ensuring better coverage of the region of interest.

With slight variations depending on the region of parameter space, \cite{Sakon:2022ibh} reports that $10\%$ of BBH injections have fitting factor smaller than $\sim 0.98$, while for our bank the $10^\text{th}$ percentile is around $0.975$.
Even though it is hard to compare the results directly due to different injection sets, it seems fair to state that, compared to \cite{Sakon:2022ibh}, our template bank provides slightly worse injection recovery.
On the other hand, our template bank has $30\%$ {\it less} templates, matching the number of templates placed by \texttt{sbank} in the same region, as reported by \cite{Sakon:2022ibh}. With an accurate treatment of the low mass corner, the coverage of our template bank will easily match the one of \cite{Sakon:2022ibh}, with a comparable bank's size.

\section{Novel applications of the method} \label{sec:novel_applications}

Our template placement method allows for several exciting applications in GW data analysis.
Obviously, the most straightforward application is the generation of high-dimensional template banks, such as a precessing and/or HM banks. While in principle it is possible to generate these high-dimensional banks with a stochastic placement method, very few of such banks have been generated so far, mostly due to the enormous computational cost of choosing the right parameter space and of computing the match between templates. Their generation becomes feasible thanks to \texttt{mbank}.

Besides efficient high-dimensional bank generation, our method can be used for other purposes as well. These include choosing the appropriate parameter space to cover by forecasting the size of a bank or selecting the appropriate coordinates to cover a given region of binary systems. Moreover, our normalizing flow could be used as a proposal for a stochastic placement algorithm or to create datasets for machine-learning applications in GW data analysis.

In what follows, we generate a large precessing template bank and a large aligned-spin HM bank. Additionally, we provide a detailed discussion of other innovative applications of our code.

\subsection{A precessing bank} \label{sec:precessing_bank}

\begin{figure*}[t]
		\includegraphics[scale = 1.]{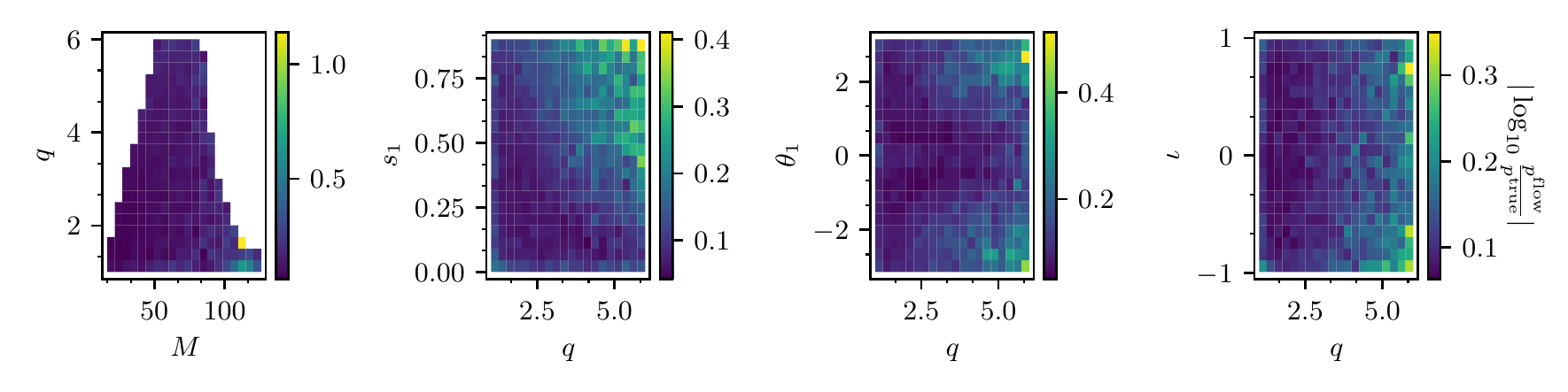}
		\caption{Accuracy of the normalizing flow trained used to generate the precessing bank in Sec.~\ref{sec:precessing_bank}. The accuracy is expressed in terms of the logarithmic ratio between the template density PDF $p^\text{true}$ Eq.~\eqref{eq:pdf_uniform} and its approximation $p^\text{flow}$ given by the flow. The flow accuracy is evaluated on $40000$ test points.}
		\label{fig:precessing_flow}
\end{figure*}

\begin{figure*}[t]
		\includegraphics[scale = 1.]{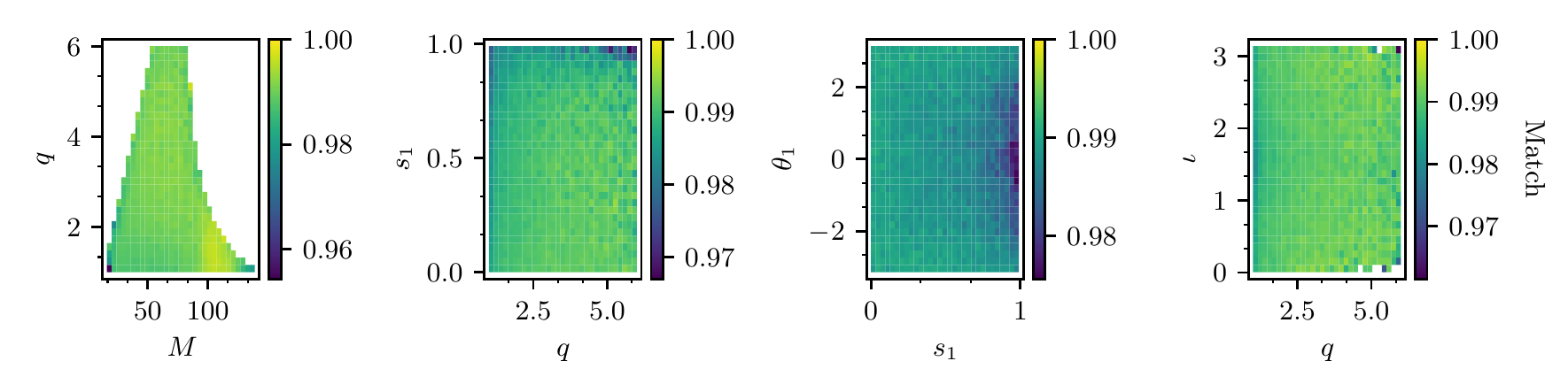}
		\caption{Fitting factor study of the precessing bank, introduced in Sec.~\ref{sec:precessing_bank}. For each bin, we color-code the median fitting factor of $10^5$ injections sampled ``On manifold", as described in the text.}
		\label{fig:precessing_fitting_factor}
\end{figure*}

\begin{figure}[t]
	\centering
	\includegraphics[scale = 1.]{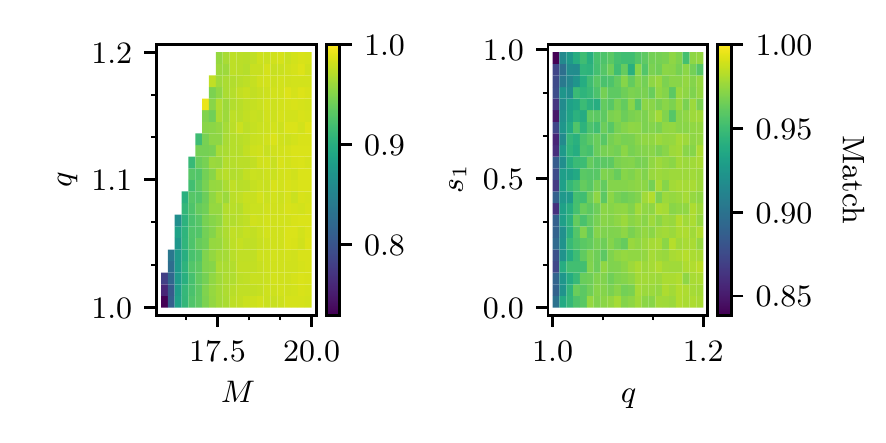}
	\caption{Fitting factor study of the precessing bank of Sec.~\ref{sec:precessing_bank}. Unlike Fig.~\ref{fig:precessing_fitting_factor}, here we focus on the low $q$, low $M$ region, where the random placement method fails. For each bin, we color-code the median fitting factor of $5 \times 10^4$ injections sampled ``On manifold", as described in the text.}
	\label{fig:precessing_fitting_factor_zoom}
\end{figure}

\subsubsection{Choosing the parameter space}

The main difficulty in generating a precessing bank lies in the huge size of the parameter space. As we show below, a precessing bank can easily have {\it billions} of templates, even when covering the mass range routinely explored by ``standard" searches. As current search pipelines can handle only up to a few {\it million} templates, due to computational cost limitations, the size of a bank sets very stringent constraints in the selection of a suitable parameter space to explore with a GW search.

Another difficulty, related to the first, arises from the choice of the BBH coordinates to include in the bank, i.e. the choice of manifold.
In principle, a precessing BBH system is described by $10$ parameters (two masses, six spins, and two angles).
However, not all of them are important, as large changes in some parameters do not result in large changes in the waveform morphology. Thus, including them in the bank does not yield any obvious improvement and, on the contrary, it may lead to vanishing metric eigenvalues, which would degrade the metric predicitivity, hence the template placement. The latter point is discussed with more details in Sec.~\ref{sec:other_applications}.

Finally, a more technical complexity arises from the fact that in high dimensional spaces, both the training of a normalizing flow (see Sec.~\ref{sec:flow_validation}) and the template placement become harder, hence possibly harming the quality of the template bank.

All these difficulties imply that great care must be taken when deciding both the parameter space and the BBH variables to include in the bank.
The choices are entangled, since covering different manifolds with the same mass range can produce banks of very different sizes.
Roughly speaking, choosing a lower dimensional sub-manifold reduces the bank size, at the cost of a loss in the bank's ability to cover the high dimensional space.

To choose a manifold, we rely on the theory. In \cite{Schmidt:2014iyl}, the authors find that the effect of the four in-plane spin components (i.e. $s_\text{1x}, s_\text{1y}, s_\text{2x}, s_\text{2y}$) can be well approximated by a single precessing spin parameter $\chi_P$ assigned to the $x$-component of the heavier object's spin.
Thus, a generic precessing system is roughly equivalent to a system with
\begin{align*}
	\mathbf{s}_\text{1} &= (\chi_P, 0, s_\text{1z}) \\
	\mathbf{s}_\text{2} &= (0, 0, s_\text{2z})
\end{align*}
effectively creating an explicit mapping between a six dimensional spin manifold to a three dimensional one.
In a later work \cite{Thomas:2020uqj}, it is suggested that to capture the combined effect of precession and HM, a two-dimensional spin parameter $\vec{\chi_P}$ is needed. In this case, the mapping is between  a six-dimensional spin manifold to a four-dimensional one.

Both works suggest that the in-plane components of the spin on the lighter object (i.e. $s_\text{2x}, s_\text{2y}$) can be neglected, reducing the dimensionality of the parameter space.
Moreover, since we are not currently concerned with precession combined with HM\footnote{In such a space, the template banks would be unfeasibly large!}, we can rely on the one-dimensional effective spin mapping \cite{Schmidt:2014iyl} to also neglect the $y$-component of the spin of the heavier object, $s_\text{1y}$.

We then consider only three out of six spin components, $s_\text{1x}, s_\text{1z}$ and $s_\text{2z}$, where all the effects of precession are included in $s_\text{1x}$. To obtain accurate coverage, we also need to include the inclination $\iota$ in the manifold.
Some investigations showed that the inclusion of the reference phase $\varphi$ yields a (almost) degenerate metric, which, by dramatically undercovering the space, negatively affects the placement.
Luckily, as injection studies show that neglecting $\varphi$ does not harm the bank's effectiveness, we can exclude $\varphi$ from the set of parameters. However, this might not be the case if we include both precession and HMs.

\begin{figure}[t]
	\centering
	\includegraphics[scale = 1.]{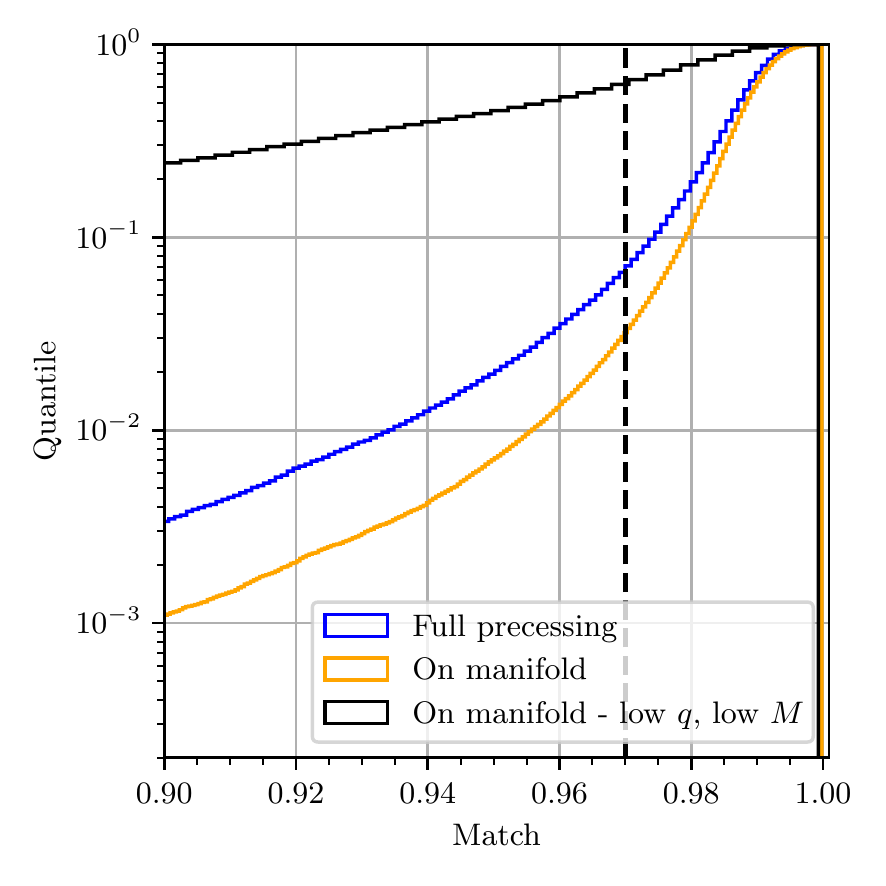}
	\caption{Cumulative fitting factor for the precessing bank introduced in Sec.~\ref{sec:precessing_bank}. The $10^5$ injections ``Full precessing" have isotropic spins, while the $3\times 10^5$ precessing injections ``On manifold" are sampled on the manifold \texttt{logMq\_s1xz\_s2z\_iota} and they have $s_\text{1y} = s_\text{2x} = s_\text{2y} = \varphi = 0$.
	For the injections ``On manifold", we plot separately the low $q$, low $M$ corner, characterized by $q \leq 1.2$ and $M \leq \SI{20}{\mathrm{M_\odot}}$. The other two histograms exclude this region. }
	\label{fig:precessing_hist}
\end{figure}

To summarize, we find that the $6$ variables $M$, $q$, $s_\text{1x}$, $s_\text{1z}$, $s_\text{2z}$ and $\iota$ provide a sufficiently complete description of waveforms in the precessing space.
This claim is confirmed by an injection study presented in Fig.~\ref{fig:precessing_hist}, where we see that more than $93\%$ of the injections covering the $10$ dimensional precessing space have a fitting factor greater than the minimal match target of $0.97$.
We note that a precessing template bank with HMs will likely need to sample two additional variables $s_\text{1y}$ and $\varphi$, hence increasing the dimensionality to $8$ \cite{Thomas:2020uqj}.

Regarding the search parameter space, we are interested to target BBHs where precession is stronger as such systems are most likely to be missed by current searches \cite{PhysRevD.102.041302, Fairhurst:2019vut}.
Precession is more visible for high mass ratio, edge-on\footnote{An edge-on system is observed with inclination $\iota\simeq \pi/2$.} systems and for high values of spins \cite{CalderonBustillo:2016rlt}. Moreover, as more cycles are detectable, precession effects will be stronger for longer signals due to the accumulation of the phasing effects of precession. These considerations suggest that very asymmetric, low mass systems, such as the neutron star-black hole (NSBH) space, would be an ideal target for a precessing bank. However, as shown below in Sec.~\ref{sec:other_applications} searching the full NSBH region is unfeasible, as hundreds of millions of templates would be needed.

For this reason, we restrict ourselves to a different, less extreme, region of the parameter space. After several investigations, made possible by the speed and flexibility of our approach, we found that a parameter space with component masses in the range $[8, 70]\,\mathrm{M_\odot}$, with a mass ratio cut-off of $6$, produces a bank with a manageable size.
In this space, we obtain a precessing bank with $\sim 2 \text{ millions}$ templates. Extending the parameter space to lower masses (or higher mass ratios) results in much larger banks, pushing the limits of current pipelines.

In closing, we stress again that the investigations above are made possible by \texttt{mbank}, since they rely on fast template bank generation across a variety of manifolds and ranges of coordinates.


\begin{figure}[t]
	\centering
	\includegraphics[scale = 1.]{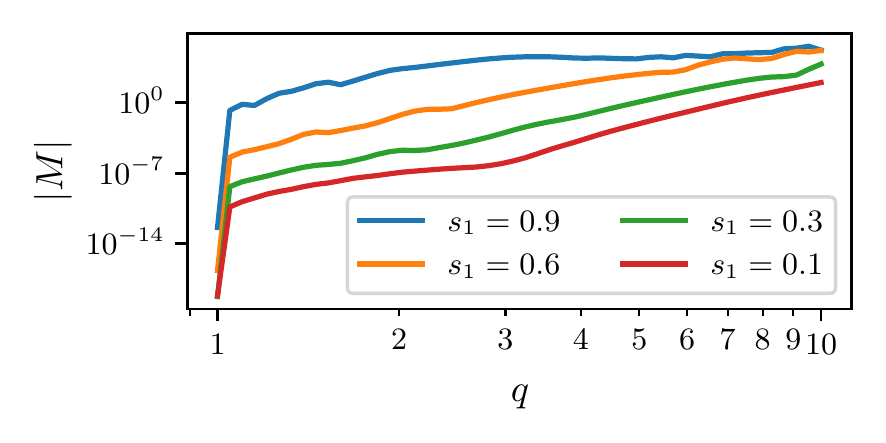}
	\caption{Determinant of the metric $|M|$ as a function of mass ration $q$ for different values of $s_1$. The metric is evaluated on the manifold \texttt{Mq\_s1xz\_s2z\_iota}, with $M = 10\mathrm{M_\odot}$, $\theta_1 = \pi/2$, $s_\text{2z} = -0.3$ and $\iota = \pi/2$. It is manifest that in all cases, the metric determinant vanishes while $q \to 1$.}
	\label{fig:metric_vs_q}
\end{figure}

\subsubsection{Generating and validating the bank}

As stated above, our precessing bank covers the manifold \texttt{logMq\_s1xz\_s2z\_iota}, with coordinates $\log_{10}M$, $q$, $s_\text{1}$, $\theta_\text{1}$, $s_\text{2z}$ and $\iota$.
We consider BBHs with individual masses between  $8$ and $70\,\mathrm{M_\odot}$, with a maximum mass ratio $q = 6$.
The other variables $s_\text{1}$, $\theta_\text{1}$, $s_\text{2z}$ and $\iota$ cover the set $[0, 0.9]\times[-\pi, \pi]\times[-0.99, 0.99]\times[0, \pi]$.

To compute the metric, we use the Advanced LIGO O4 sensitivity estimate \cite{O4_PSDs} and we set a frequency range of $[15, 1024] \,\SI{}{Hz}$, employing the approximant \texttt{IMRPhenomXP} \cite{Pratten:2020ceb}.
We train a normalizing flow with $3$ layers with $100$, $100$ and $60$ hidden features respectively, using a dataset of $4\times 10^5$ points.
The flow performance after training is reported in Fig.~\ref{fig:precessing_flow}.
To generate the bank, we use a minimal match requirement of $0.97$, with a covering fraction $\eta = 0.95$, estimated with $3000$ livepoints.
In a similar way to what was done for the ``All-sky" template bank, we also train a normalizing flow to target the high total mass region with $M>\SI{100}{\mathrm{M_\odot}}$. We use the latter to place templates with the same covering fraction $\eta = 0.95$, with great benefits.
The overall bank has $1605625$ templates, plotted in Fig.~\ref{fig:corner_precessing}.

This bank generation took a few hours in total: ${\sim \SI{1}{hour}}$ for the dataset generation, ${\sim \SI{30}{minutes}}$ for the training of the flow and ${\sim \SI{5}{minutes}}$ for the template placing.
All the steps above runned on a single core, using less than $\SI{4}{GB}$ of memory.
We highlight that our time and memory requirements are a fraction of those of a similar bank with the state-of-the-art stochastic algorithm.

The template distribution reported in Fig.~\ref{fig:corner_precessing} shows a spike in the template density for $\theta_1 = \pm\pi$ (close to the non-precessing limit) in the high mass ratio and high $s_1$ region. Some investigations indicate that these are not artefacts introduced by the normalizing flow.
Whether the feature is physical or is due to the behaviour of the waveform approximant in the non-precessing limit remains an open question which needs more inspection.

To study the performance of our template bank, we generate two injections sets, with masses sampled uniformly in $\log m_1$ and $\log m_2$.
The first set, labeled ``Full precessing" has fully precessing injections (with two $3D$ spins and varying $\varphi$). The second one, denoted as ``On manifold", has injections lying on the manifold \texttt{logMq\_s1xz\_s2z\_iota}, hence covering a subset of the ``Full precessing" set.
The latter set is needed to asses the coverage of the bank on the manifold on which the templates lie and thus is a measure of the templates' placement accuracy.
On the other hand, the ``Full precessing" injection set evaluates the ability of the bank to recover a generic precessing signal, hence assessing the quality of our choice of manifold.
Clearly, this is the injection set that is most relevant for designing the bank for a fully precessing search.

We report the results of our study in Fig~\ref{fig:precessing_hist}, in the form of a histogram of the fitting factors, and in Fig.~\ref{fig:precessing_fitting_factor}, where we study the dependency of the fitting factor across the parameter space.
Fig.~\ref{fig:precessing_fitting_factor_zoom} reports the same fitting factor study focused on the low $q$, low $M$ region.

As is clear from Fig.~\ref{fig:precessing_fitting_factor_zoom} and~\ref{fig:precessing_hist}, the random template placement method {\it fails} for the low $q$, low $M$ region, with $q \leq 1.2$ and $M \leq \SI{20}{\mathrm{M_\odot}}$, where only $\sim 40\%$ of the injections ``On manifold" have a fitting factor higher than $0.97$.
On the other hand, outside the low $q$, low $M$ corner, the template bank provides a good coverage: $97\%$ of the injections ``On manifold" has a fitting factor large than $0.97$.

The poor performance for low mass ratio and low masses was also observed in the ``All-sky" template bank in Sec.~\ref{sec:all_sky_comparison}, although less severe.
Such failure be explained by two combined causes.
First of all, as noted above, the random method is unable to cover ``sharp" corners of the parameter space, due to the lack of appropriate boundary treatment: this can (and does) severly limit the bank's ability to cover the space.
Moreover, we observe that for $q \to 1$ the metric determinant goes rapidly to $0$, meaning that very few templates are placed.  This is shown in Fig.~\ref{fig:metric_vs_q}, where we plot $|M|$ as a function of $q$ keeping constant all the other coordinates \footnote{Although not reported here, the same behaviour is observed for ``standard" signals.}.
The two effects combines together in the low $q$, low $M$ region, which is drastically undercovered. The same issue is not observed anywhere else in the parameter space.

In principle, we could remedy the problem by extending the covered region to lower masses and higher $q$: this would make sure that the low $q$, low $M$ target region does not lie at the boundaries of the bank anymore.
However, the lack of coverage in this region is not a major concern for the bank's effectiveness in a real search scenario. Indeed, precession for $q\sim 1$ has very little effect on the BBH waveform and a precessing system with symmetric masses would likely be detected by current aligned-spin searches.

In Fig.~\ref{fig:precessing_fitting_factor}, we see that the coverage is rather uniform across the parameter space. The median fitting factor slightly drops for the high $q$ high $s_1$ corner of the parameter space.
As shown in Fig.~\ref{fig:precessing_flow}, the flow performance degrades in that undercovered corner of the space: the true template density $\sqrt{|M|}$ is underestimated by the normalizing flow, which accordingly places less templates than optimal.

The fitting factor of the ``Full precessing" injection set is fairly good, with only $7\%$ of the injections (outside the ``low $q$, low $M$" region) below the target match. This means that the $\chi_P$ approximation that motivates our choice is robust: the manifold \texttt{logMq\_s1xz\_s2z\_iota} provides a faithful low-dimensionality representation of the entire precessing parameter space.

\begin{figure*}[t]
	\includegraphics[scale = 1.]{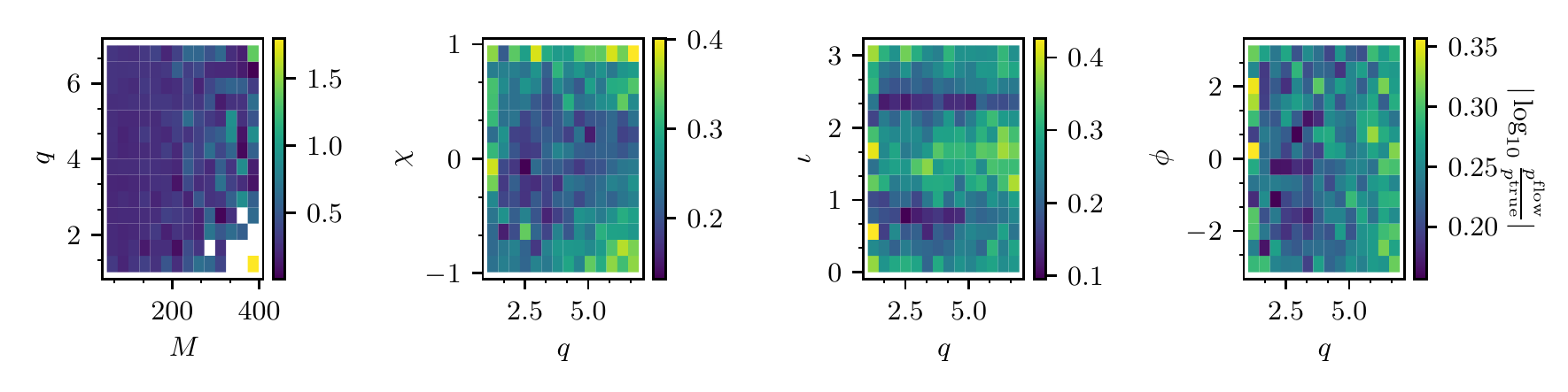}
	\caption{Accuracy of the normalizing flow trained used to generate the aligned-spin HM bank of Sec.~\ref{sec:HM_spinning_bank}. The accuracy is expressed in terms of the logarithmic ratio between the template density PDF $p^\text{true}$ Eq.~\eqref{eq:pdf_uniform} and its approximation $p^\text{flow}$ given by the flow. The flow accuracy is evaluated on $40000$ test points.}
	\label{fig:HM_flow}
\end{figure*}

\begin{figure*}[t]
	\includegraphics[scale = 1.]{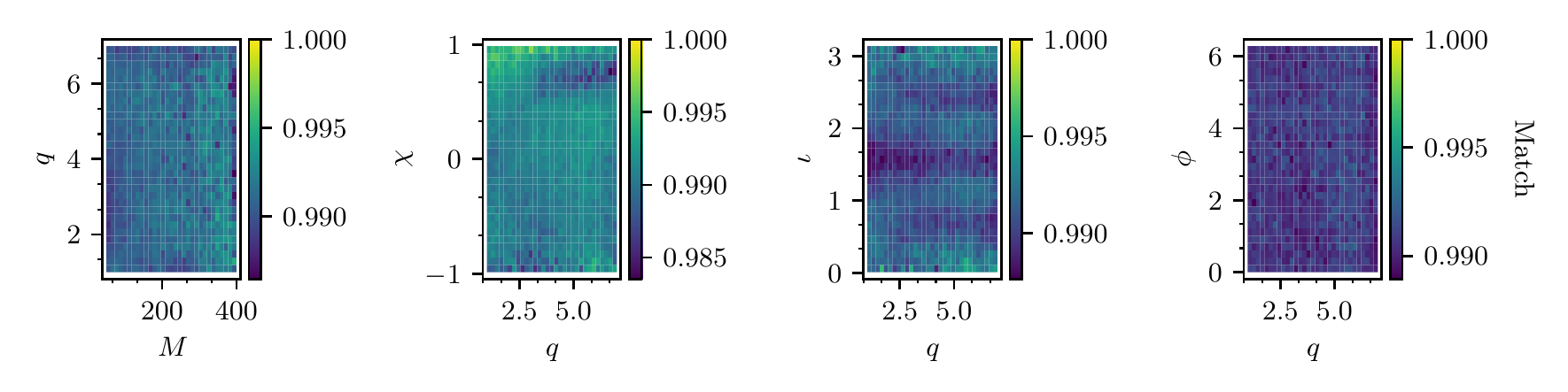}
	\caption{Fitting factor study of the aligned-spin HM bank, introduced in Sec.~\ref{sec:HM_spinning_bank}. For each bin, we color-code the median fitting factor of $10^5$ injections sampled uniformly from the the parameter space.}
	\label{fig:HM_fitting_factor}
\end{figure*}

\begin{figure}[t]
	\centering
	\includegraphics[scale = 1.]{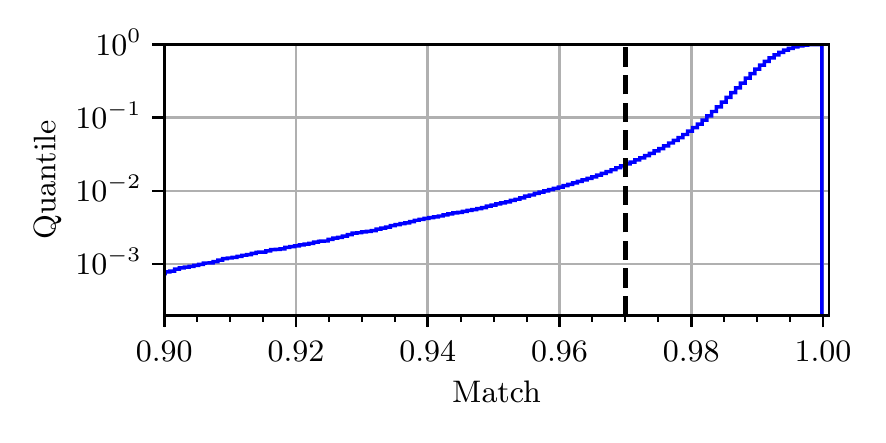}
	\caption{Cumulative fitting factor for the aligned-spin HM bank described in Sec.~\ref{sec:HM_spinning_bank}. The histogram is built upon $10^5$ injections sampled from the manifold.}
	\label{fig:HM_hist}
\end{figure}


\subsection{An aligned-spin HM bank} \label{sec:HM_spinning_bank}

In a sense, aligned-spin HM template banks are easier to generate than precessing ones, due to a smaller dimensionality of the parameter space. Indeed, a generic aligned-spin binary system with HMs is characterized by $6$ parameters (two masses, two spins and two angles $\iota$ $\varphi$) but, as for the non-HM case, the spin effects can be easily parametrized with an effective spin parameter, reducing the number of dimensions to $5$.
Note that here we deal with one dimension more than in the nonspinning HM bank produced in Sec.~\ref{sec:HM_comparison}.
Despite less uncertainties in the choice of manifold than in the precessing case, the parameter space is very large and producing a template bank of a feasible size still requires a careful choice of the region to target.

We used \texttt{mbank} to generate an HM aligned-spin bank, covering the high mass region of the BBH parameter space.
High mass events are notoriously hard to detect \cite{LIGOScientific:2021tfm, Chandra:2021wbw}. As they are very short, their morphology matches closely non-Gaussian transient noise bursts, also called {\it glitches}, \cite{Blackburn:2008ah, Zevin:2016qwy, LIGOScientific:2016gtq, LIGO:2021ppb}. In this scenario, a more realistic model for the waveform can improve the detectability of such signals, thanks to both an increase in recovered SNR and to a more accurate signal-based veto \cite{Babak:2005kv, PhysRevD.95.042001}.
Several studies \cite{Pekowsky:2012sr, Capano:2013raa, Varma:2014jxa, CalderonBustillo:2015lrt} confirmed this claim, finding that failing to consider HMs in GW searches can lead to a large sensitivity loss for large mass ratios $q\gtrsim 4$ and high masses $M \gtrsim \SI{100}{\mathrm{M_\odot}}$ \cite{CalderonBustillo:2016rlt}.

Consequently, our bank covers the manifold \texttt{logMq\_chi\_iotaphi}, sampling $\log_{10}M$, $q$ and $\chi_\text{eff}$ as well as inclination and reference phase.
We consider templates with total mass $M$ between $\SI{50}{\mathrm{M_\odot}}$ and $\SI{400}{\mathrm{M_\odot}}$ and a mass ratio smaller than $7$. The effective spin lies in range $[-0.99, 0.99]$ and, as usual, $\iota \in [0, \pi]$ and $\varphi \in [-\pi, \pi]$.
We use the Advanced LIGO O4 sensitivity estimate \cite{O4_PSDs} and we set a frequency range of $[10, 1024]\,\SI{}{Hz}$, with approximant \texttt{IMRPhenomXHM} \cite{Garcia-Quiros:2020qpx}.

We generate a dataset with $4\times 10^5$ points and train a normalizing flow with $4$ layers, each with $n_\text{hidden} = 60$ hidden features. The accuracy of the normalizing flow is reported in Fig.~\ref{fig:HM_flow}.
For the template placement, we use a minimal match requirement of $0.97$ and set a covering fraction $\eta = 0.8$, estimated with $10000$ livepoints.
The overall bank gathers $2115299$ templates, which are plotted in Fig.~\ref{fig:corner_HM_spinning}.
The bank generation took roughly the same time as for the precessing bank.

We study the bank performance with $10^5$ injections and report their fitting factor in Fig.~\ref{fig:HM_fitting_factor} and Fig.~\ref{fig:HM_hist}.
Our injection study shows that only $\sim 2\%$ of the injections have a fitting factor smaller than the target of $0.97$, with a median fitting factor of $0.99$. We can conclude that the bank provides good coverage of the parameter space.
Moreover, the fitting factor is rather constant across all the parameters space. As was also the case for the HM bank introduced in Sec.~\ref{sec:HM_comparison}, there are not regions which are undercovered by the template banks.
Also the accuracy of the normalizing flow does not vary too much over the parameter space, showing a bad performance only in the region with high total mass and low mass ratio.

We note that, in order to achieve good performance in the two HM banks presented in this work, we set a covering fraction of only $\eta = 0.8$. This is significantly lower than what we used for the non-HM banks and also lower than the recommended value of $\eta = 0.9$ in \cite{Coogan:2022qxs}.
This means that, unlike the non-HM case, the metric match in Eq.~\ref{eq:metric_definition} {\it underestimates} the ``true" match. In this scenario, the covering fraction estimated with the livepoints (which makes use of the metric) also underestimates the ``true" covering fraction. Therefore, a lower value of $\eta$ is enough to obtain an acceptable coverage. This is not the case for non-HM banks.
The reason why this happens only for HM banks is currently not understood and requires more investigation.

\subsection{Other possible applications} \label{sec:other_applications}

\begin{figure}[t]
	\centering
	\includegraphics[scale = 1.]{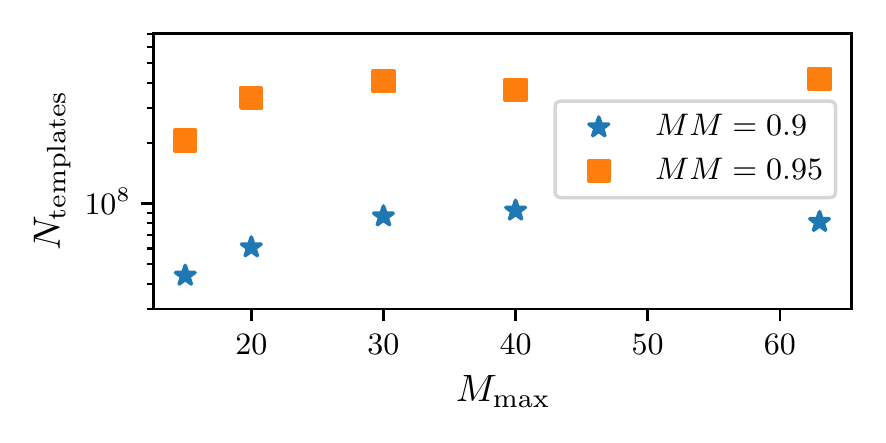}
	\caption{Study of the size of a template bank in the neutron star-black hole parameter space. Each point refers to a template bank on the manifold \texttt{logMq\_s1xz\_iota}, covering a total mass range $M\in[M_\text{min}, M_\text{max}]$. The component masses are limited to $m_1 \in [10, 60]\,\mathrm{M_\odot}$ and  $m_2 \in [1, 3]\,\mathrm{M_\odot}$, with mass ratios $q \in [3.3, 15]$. In the plot we report the number of templates $N_\text{templates}$ as a function of the maximum total mass $M_\text{max}$, for different minimal match requirements. The resulting banks are huge, with tens of millions of templates, showing that a search for precessing NSBH binaries is still prohibitively costly.}
	\label{fig:NSBH_size}
\end{figure}

The speed of the bank generation, together with the flexibility of the flow in sampling from the parameter space, allows for several novel applications of our work to GW data analysis, besides producing high-dimensional template banks.
Without being exhaustive, we discuss below some of the new possibilities.

\paragraph{Selecting the parameter space to target}\label{par:parameter_space_selection}
As already discussed, the choice of the parameter space to target in GW searches can be challenging, as it is hard to obtain a reliable forcast of the number of templates needed for accurate coverage.
Moving towards high-dimensional template banks, the number of templates increases by orders of magnitude and the standard stochastic approach suffers from memory issues due to the storage of the waveforms needed for the match calculation.
This in turn makes it difficult to even explore high-dimensional spaces, as the current algorithms time-out by the time the bank reaches several million templates

Our method has a low memory footprint and this makes possible to forecast the number of templates in a given parameter space, providing invaluable information to choose an appropriate target for the search.
To do so, the interested user might train a normalizing flow on a large region of the parameter space and then place templates in a subregion, without the need to store them. Sampling in a subregion can be easily completed with the use of rejection sampling.

A natural candidate to demonstrate the usefulness of this technique is the precessing NSBH parameter space.
Indeed, due to the large mass asymmetry of NSBH systems (i.e. high $q$), precession has a strong imprint on the waveform, leading to a very large volume to cover by a template bank.
To study the number of templates needed to cover the space, we train a normalizing flow model on the manifold \texttt{logMq\_s1xz\_iota}\footnote{We neglect any spin on the lighter object, a neutron star. This is physically motivated by the fact that a NS is expected to have low or no spins.} for systems with masses $m_1 \in [10, 60]\,\mathrm{M_\odot}$ and  $m_2 \in [1, 3]\,\mathrm{M_\odot}$, with mass ratios $q \in [3.3, 15]$.
The other coordinates $s_\text{1}$, $\theta_\text{1}$ and $\iota$ vary in set $[0, 0.9]\times[-\pi, \pi]\times[0, \pi]$.
As above, we use the approximant \texttt{IMRPhenomXP}, in a frequency range of $[15, 1024]\,\SI{}{Hz}$.

To study the parameter space size, we run our template placement algorithm for varying maximum total mass $M_\text{max}$ and we measure the number of templates needed to achieve a covering fraction of $\eta = 0.9$ for different minimal match requirements.
Since we do not store and validate the template banks, there is no guarantee that the resulting banks provide a satisfactory coverage. The procedure is just meant to obtain an order of magnitude estimation of the bank size.

As shown in Fig.~\ref{fig:NSBH_size}, the precessing NSBH parameter space is huge. With a minimal match requirement of $0.9$, around $100$ million templates are needed to cover the full space. Around half of the templates are in the low total mass region with ${M \in [11,15]\,\mathrm{M_\odot}}$. The numbers agree with the investigations carried out in \cite{McIsaac:2023ijd}.
To cover the space with a minimal match of $0.95$, around five times more templates are needed.

The magnitude of the precessing NSBH space makes it nearly impossible to use traditional matched filtering techniques to search for such signals. It thus becomes compelling to either develop new search techniques \cite{McIsaac:2023ijd} or to improve the computational power available.

Thanks to our method, similar estimates can easily be done for other regions of the BBH parameter space (e.g. targeting eccentric BBHs), thus providing invaluable information to plan future high-dimensional GW searches.

\paragraph{Manifold selection}\label{par:manifold_selection}

The metric eigenvalues and eigenvectors can give an interesting piece of information about the relative importance of the coordinates of the manifold.
Let $\lambda_i$ and $\mathbf{v}_i$ be the i-th eigenvalue and eigenvector respectively of the metric $M_{ij}$.
We can think of each eigenvector $\lambda_i$ as a measure of the relative importance of the eigenvector $\mathbf{v}_i$, which represents a linear combination of the coordinates.
We can then introduce the following quantity for each coordinate $j$, which we call {\it coordinate importance}:
\begin{equation}
\mathcal{I}_j = \left| \sum_i \lambda_i (\mathbf{v}_i)_j \right|
\end{equation}
where $(\mathbf{v}_i)_j$ is the j-th component of the i-th eigenvector.
It is a weighted average over the projection of each eigenvector along a given coordinate. Heuristically, an ``important" coordinate will give a larger contribution to the ``important" eigenvectors (i.e. with larger eigenvalues).

This quantity might be used to create a hierarchy among the coordinates and, when choosing the manifold to cover, it can offer a useful criteria to decide which quantities to include in the bank.
For example, in the manifold \texttt{logMq\_s1xyz\_s2z\_iotaphi}, the variable $\log_{10}M$ has an importance of $5\times 10^4$, while variables $q, s_1, \theta_1$ and $s_\text{2z}$ have importance two orders of magnitude less. This implies that a template bank must include (besides the total mass) all the variables $q, s_1, \theta_1$ and $s_\text{2z}$.
On the other hand, coordinates $\phi_1$ (controlling the magnitude of $s_\text{1y}$) and the angles $\iota$ and $\varphi$ have an ``importance" of one order of magnitude less than all the other quantities. As a consequence, the latter three play a smaller role in covering the space and they can be possibly ignored (or perhaps only one of them can be included).

Of course, this line of reasoning is heuristic and whether a manifold is suitable or not to cover the space must be checked by means of an injection study. However, the study of the relative importance between coordinates can give an educated guess on the manifold to cover and serve as a starting point for the trial and error process of manifold selection.

\paragraph{A proposal for the stochastic template placement}

Our normalizing flow finds an obvious application within a stochastic placement algorithm. According to the stochastic algorithm, template proposals are randomly drawn from an analytical PDF, which is specifically design to approximate Eq.~\eqref{eq:pdf_uniform} in the non-spinning case.
A good proposal is crucial to reduce the template rejection rate, hence reducing the overall run time.

The normalizing flow is a natural candidate for a proposal distribution, since it goes beyond the non-spinning BBH approximation, allowing for more physics to be considered. Implementing a normalizing flow within the stochastic algorithm will most likely provide a computational benefit, due to a more efficient proposal.

\paragraph{Generating datasets for machine learning applications}

The recent years have seen a burst of machine learning application to GW data analysis, covering all fields of the analysis of compact binary systems from waveform modelling \cite{Khan:2020fso, Schmidt:2020yuu, Thomas:2022rmc, Tissino:2022thn} to GW searches \cite{Gebhard:2019ldz, Schafer:2020kor, Schafer:2021fea, Baltus:2021nme} and parameter estimation \cite{Green:2020hst, Alvares:2020bjg, Williams:2021qyt, Langendorff:2022fzq, Williams:2023ppp}.

For all these applications, it is crucial to have high quality datasets of waveforms for training purposes.
The goodness and the applicability of the model strongly relies on the distribution of waveforms in the dataset and substantial time is often spent in tuning the dataset composition to achieve optimal performance.
The waveforms in such datasets can be sampled using our normalizing flow model, thus covering the space accurately. In many cases this may prove beneficial.

\section{Future prospects} \label{sec:improvements}

Clearly, our work can be improved and expanded in several directions. In this section, we discuss some possible advancements.

\paragraph{Introducing a new metric}

As shown in App.~\ref{app:metric_definition}, the Hessian of the match (with which we identify the metric) does not always approximate the behaviour of the true match in a neighbourhood of a point.
For instance, on the manifold \texttt{Mq\_s1xyz}, consider the ellipse $\mathcal{E}_0$, centered on ${\theta_0 = (\SI{10}{\mathrm{M_\odot}}, 7, 0.6, 2, 2)}$ of all the points $\theta$ with {\it metric} match with the center higher than $0.97$.
It turns out that only $\sim 50\%$ of the points inside $\mathcal{E}_0$ have a match higher than $0.97$. The situation gets worse for smaller mass ratio, when the metric determinant vanishes, and it can significantly vary among different manifolds.

While this hasn't affected (too much) the effectualness of our template bank, the failure of the metric approximation is concerning and can negatively influence the placement, especially in presence of a parameter with a small impact on the waveform. The interested reader is encouraged to read App.~\ref{app:metric_definition}.

\paragraph{Exploring different flow architectures}

In this work, we only considered Masked Autoregressive Layers for our normalizing flow architecture. Of course, other choices are available in the literature and could possibly improve the flow accuracy. Further work should implement some of these and assess the (possible) gain in accuracy.
Possible transformations include coupling layers \cite{Dinh2014NICENI,Dinh2016DensityEU} or residual flows \cite{NIPS1999_e6384711, Behrmann2018InvertibleRN}.

As discussed in Sec.~\ref{sec:normalizing_flow}, it is very beneficial to use a transformation like Eq.~\eqref{eq:first_transform} as the first layer of the normalizing flow. Future work can find a different transformation offering better performance.

\paragraph{Estimating the covering fraction with importance sampling}

An accurate evaluation of the covering fraction in Eq.~\eqref{eq:coverage} is crucial to providing a realistic estimation of the template number and hence good coverage.
Currently we estimate the covering fraction by using the approximation to the volume element given by the normalizing flow.
We can increase the accuracy by computing the integral in  Eq.~\eqref{eq:coverage} with importance sampling:
\begin{equation}\label{eq:coverage_estimate_IS}
	\hat{\eta}(\mathcal{T}) \simeq \frac{1}{\sum_i w_i} \sum_i c(\theta_i) w_i 
\end{equation}
where the livepoints are sampled from the flow and are weighted with weights $w_i = \frac{\sqrt{|M(\theta_i)|}}{p^\text{flow}(\theta_i)}$.
The weights make sure that we evaluate the unapproximated version of the integral, i.e. using the true volume element and not its flow approximation.

In a practical application, it is wise to prevent the weights to grow indefinitely, as this can negatively impact the estimation of the covering fraction. For this reason, we clip the weights to a maximum value of $W$: ${w_i = \min\left(\frac{\sqrt{|M(\theta_i)|}}{p^\text{flow}(\theta_i)}, W_\text{max} \right)}$. The tuning of $W_\text{max}$ deserves more attention, as it can really impact the bank performance.

Some tests have shown that importance sampling delivers larger banks, thus with better coverage but with an increased variance in the number of templates. However, in some occasions, one or a few livepoints can dominate the sum (i.e. have very large weight), making the covering fraction computation less robust in case of flow inaccuracies.
More work is required to treat such cases and successfully implement this new feature.


\paragraph{Exploring different placement methods}

While the random template placement method in use has proven its efficacy, other alternatives are certainly possible. A different placement method is appealing to reduce the bank size without degrading its performance, as random template banks tend to place more templates than needed.

First, one could use the metric to reject templates that are too close to each other. This would be a variation of the stochastic algorithm, where distances are computed with the metric and not with the true match. While this may prove unfeasibly slow in some cases, it can still be computationally more efficient than with the brute force match computation. As a compromise, a random template bank with low covering fraction and minimal match might be given as starting point for the iteration (i.e. a seed bank).

One could also devise alternative strategies to sample from the flow latent space, such as using quasi Montecarlo sampling or even setting points on a lattice.
Since the coordinates of the templates will be correlated with each other, we cannot compute iteratively the covering fraction as described in Sec.~\ref{sec:template_placing}. For this reason the suitable bank size needs to be computed with other methods, before selecting the templates.

Regardless of the placement method, the templates in a bank may still not be placed optimally, creating over(under)-dense regions. This is especially true for the {\it random} method used here. For this reason, it may be beneficial to add a post-processing step to move or remove some templates \cite{Indik:2017vqq}.

\paragraph{Encoding the metric into the flow?}

A fascinating path to explore is to encode information about the metric $M_{ij}$ inside the flow transformation.
So far, the normalizing flow $\phi_W$ is trained in such a way that the determinant of the Jacobian $\det J_{\phi_W}$ matches the determinant of the metric.
Thus, among the $\frac{D(D-1)}{2}$ free components of $J_{\phi_W}$, only one of them is constrained during the training. This leaves a lot of degeneracy in $J_{\phi_W}$.
One could break such degeneracy by imposing the additional constraint that the Jacobian of the flow matches the metric $M_{ij}$:
\begin{equation}
	(J_{\phi_W})_{ij} \simeq M_{ij}.
\end{equation}
Such constraint should be imposed by introducing a suitable loss function.
The approach would involve a much harder optimization problem and it remains to be assessed whether the flow has enough representation power to solve such problem.

A flow trained in this way would create an isometry (i.e. distance preserving transformation) between the latent space and the physical space. According to differential geometry, this is not possible, unless the $M_{ij}$ has zero curvature, which is not the case in general. A possible way out could be to embed the manifold of signals in an higher dimensional flat manifold, which would guarantee the existence of a solution.

As outlined, there are many open questions and issues to solve, which require significant work. The reward however would be huge: the flow would parameterize a distance preserving (and not only volume preserving) transformation, which can be used for high dimensional fast stochastic placement or even geometric placement - the holy grail of bank generation.

\section{Final remarks} \label{sec:conclusion}

We present a novel method to generate template banks covering a high-dimensional manifold of (possibly) precessing/HM/eccentric BBH signals.

Key to our method is the metric $M_{ij}$ and the derived volume element $\sqrt{|M|}$. The latter defines the number of templates that should cover an infinitesimal volume and can be seen as a probability measure on the space.
We derive here for the first time an expression for the metric suitable for precessing and/or HM signals (see App.~\ref{app:metric}). The metric is written in terms of the gradients of the waveform polarizations and is numerically stable.

To sample the templates, we introduce a novel normalizing flow model, which serves the twofold purpose of sampling from the space and providing a fast-to-compute approximation to $\sqrt{|M|}$.
Once we are able to sample from the space, we place templates using the {\it random} algorithm, which is fast and suitable to cover high-dimensional spaces.
This comes at the price of a larger bank than would be produced with the state-of-the-art stochastic algorithm, although the over-coverage becomes less severe as the number of dimensions, and correspondingly the overall size of the bank, increases.

We validate our code by evaluating the normalizing flow accuracy and the robustness of the random placement.
Moreover, with a few hours of computation, we were able to reproduce two template banks existing in the literature obtained with independent codes - a nonspinning HM bank \cite{Harry:2017weg} and an aligned-spin bank \cite{Sakon:2022ibh}.

To demonstrate the capabilities of our code, we generate two large template banks covering systems for which no or little searches have been performed: a precessing bank gathering $1.6$ million templates (Sec.~\ref{sec:precessing_bank}) and an aligned-spin HM bank formed by $2.1$ million templates (Sec.~\ref{sec:HM_spinning_bank}). We show that the two banks satisfactorily cover the space. They were both produced in a matter of hours, with minimal CPU and memory usage.
We also discuss other possible applications of our method, including the optimization of the template proposal of the stochastic algorithm, the selection of a suitable parameter space for a GW search and the generation of datasets of waveforms for the training of machine learning models.

Our code is publicly available as a package \texttt{mbank}\cite{mbank} and comes with a large number of tools to simplify the bank generation and validation.

As a final remark, we stress that our work will enable the GW community to run searches on novel regions of the BBH parameter space. Being able to generate a high dimensional bank in a few hours, the computational cost of searching new regions of the parameter space will be dominated by the actual cost of the analysis rather than the cost of prior steps.
This will allow for optimal resource allocation to search for signatures of precession, eccentricity and/or HMs, hopefully leading to exciting physics discoveries.

        \begin{acknowledgments}
		We thank Melissa Lopez Portilla, Harsh Narola, Aaron Zimmerman and Keith Riles for their precious comments. We should not forget to thank the anonymous referee who stimulated huge improvements to our work with their interesting comments.
		S.S., B.G., and S.C. are supported by the research program of the Netherlands Organization for Scientific Research (NWO). S.C is also supported by the National Science Foundation under Grant No. PHY-2309332.
		The authors are grateful for computational resources provided by the LIGO Laboratory and supported by the National Science Foundation Grants No. PHY-0757058 and No. PHY-0823459. This material is based upon work supported by NSF’s LIGO Laboratory which is a major facility fully funded by the National Science Foundation.
        \end{acknowledgments}

\appendix

\section{Details of the metric computation}\label{app:metric}

In this Appendix we report the details of the derivation of Eq.~\eqref{eq:metric_expression}, as well as the computation of the Hessian $H$ of the overlap in Eq.~\eqref{eq:overlap} in terms of the gradients of the waveform $h(\theta)$. 
In what follows, we define $\rescalar{h_1}{h_2}$ and $\imscalar{h_1}{h_2}$ to be the real and imaginary part, respectively, of $\scalar{h_1}{h_2}$.

We begin by expanding the quantity $\mathcal{M}(\theta+\Delta\theta,\theta )$ for $\Delta\theta$ around $0$. Since $\mathcal{M}(\theta+\Delta\theta,\theta )$ has a maximum for $\Delta\theta = 0$, the leading term is quadratic in $\Delta\theta$.
We obtain:
\begin{align} \label{eq:metric_derivation}
	&\mathcal{M}(\theta+\Delta\theta,\theta ) = \max_{\Delta t} \mathcal{O}(\theta + \Delta\theta, \theta, \Delta t) \nonumber\\
	& =	\max_{\Delta t} \left\{ 1+ \frac{1}{2}\left[ \partial_{ij}\mathcal{O} \Delta\theta_i \Delta\theta_j + 2  \partial_{it}\mathcal{O} \Delta\theta_i \Delta t + \partial_{tt}\mathcal{O} (\Delta t)^2 \right] \right\}  \nonumber \\
	&= 1 + \frac{1}{2}\left[ \partial_{ij}\mathcal{O} - \frac{\partial_{it}\mathcal{O} \partial_{jt}\mathcal{O}}{\partial_{tt}\mathcal{O}}\right] \Delta\theta_i \Delta\theta_j
\end{align}
where all the derivatives are evaluated at ${\Delta\theta = \Delta t = 0}$ and the explicit time maximization yields
${\Delta t = -\frac{\partial_{it}\mathcal{O} \Delta\theta_i}{\partial_{tt}\mathcal{O}}}$.

From Eq.~\eqref{eq:metric_derivation}, we can read the expression for the metric in Eq.~\eqref{eq:metric_expression} recognizing in the derivatives $\partial\partial\mathcal{O}|_{\Delta\theta, \Delta t = 0}$ the components of the Hessian matrix $H$ of the overlap.

We now compute the Hessian $H$ of the overlap in terms of the gradients of the {\it normalized} waveforms. For notational convenience, we set $h_+(\theta_1)e^{ift} = s$, we drop any dependence on $\theta_2$ and we understand $\mu = {i, t}$.
We have:
\begin{align}\label{eq:overlap_grads}
	\partial_{\mu} \mathcal{O} &= \frac{1}{\mathcal{O}} \frac{1}{1-\hat{h}^2_{+\times}}
	\left[
	\rescalar{\partial_\mu\hat{s}}{\hat{h}_+}\rescalar{\hat{s}}{\hat{h}_+} 
	+ \rescalar{\partial_\mu\hat{s}}{\hat{h}_\times}\rescalar{\hat{s}}{\hat{h}_\times} \right. \nonumber \\
	&\left. - \rescalar{\partial_\mu\hat{s}}{\hat{h}_+}\rescalar{\hat{s}}{\hat{h}_\times}h_{+\times}
	- \rescalar{\partial_\mu\hat{s}}{\hat{h}_\times}\rescalar{\hat{s}}{\hat{h}_+}h_{+\times}
	\right]
\end{align}
Differentiating another time, after some rearrangements, we get:
\begin{align}
H_{tt} &= - \rescalar{\hat{h}_+}{\hat{h}_+f^2}
			+ \frac{1}{1-\hat{h}^2_{+\times}} \imscalar{\hat{h}_\times}{\hat{h}_+f}^2 \label{eq:H_tt}\\
H_{ti} &= \imscalar{\hat{h}_+}{\partial_i \hat{h}_+f}
			- \frac{1}{1-\hat{h}^2_{+\times}} \rescalar{\hat{h}_\times}{\partial_i\hat{h}_+} \imscalar{\hat{h}_\times}{\hat{h}_+f} \label{eq:H_ti}\\
H_{ij} &= \rescalar{\hat{h}_+}{\partial_i\partial_j\hat{h}_+}
			+ \frac{1}{1-\hat{h}^2_{+\times}} \rescalar{\hat{h}_\times}{\partial_i\hat{h}_+} \rescalar{\hat{h}_\times}{\partial_j\hat{h}_+} \label{eq:H_ij}
\end{align}

To move further, we express the normalized waveform derivatives in terms of the un-normalized ones:
\begin{align*}
	\bullet&\quad \partial_i \scalar{h}{h} = \scalar{\partial_i h}{h}+ \scalar{h}{\partial_i h} = 2 \rescalar{h}{\partial_i h} \\
	\bullet&\quad \partial_i \hat{h} =\frac{1}{\rescalar{h}{h}^{3/2}} \left[ \rescalar{h}{h}\partial_i h -  \rescalar{h}{\partial_i h} h \right]
	\\
	\bullet &\quad \partial_t \hat{h} = i f \hat{h} = i f \frac{h}{\rescalar{h}{h}^{1/2}} \\
	\bullet &\quad \partial_i \partial_j \hat{h} = \frac{1}{\rescalar{h}{h}^{1/2}} \partial_{ij}h 	+3 \frac{1}{\rescalar{h}{h}^{5/2}} \rescalar{h}{\partial_i h}\rescalar{h}{\partial_j h}h \\
	&- \frac{1}{\rescalar{h}{h}^{3/2}} \left[\rescalar{h}{ \partial_{ij} h} h + \rescalar{\partial_i h}{\partial_j h}  h
		+2\rescalar{h}{\partial_{(i} h} \partial_{j)} h \right]
\end{align*}
where $A_{(ij)} = \frac{1}{2}(A_{ij}+A_{ji})$ denotes symmetrization.

Plugging this into the equations~\eqref{eq:H_tt}-\eqref{eq:H_ij}, we get:
\begin{widetext}
\begin{align}
	H_{tt} &= - \frac{1}{h_{++}} \rescalar{h_+}{f^2 {h_+}}
		+ \frac{1}{1-\hat{h}^2_{+\times}} \frac{1}{h_{++}h_{\times\times}} \imscalar{{h_\times}}{f{h_+}}^2 \label{eq:H_tt_grad} \\
	H_{ti} &= - \frac{1}{h_{++}} \rescalar{h_+}{f \partial_i h_+}
		- \frac{1}{1-\hat{h}^2_{+\times}} \frac{1}{h_{++}h_{\times\times}} \imscalar{{h_\times}}{f{h_+}} \rescalar{{h_\times}}{\partial_i h_+}
		+ \frac{\hat{h}_{+\times}}{1-\hat{h}^2_{+\times}} \frac{1}{h^{3/2}_{++}h^{1/2}_{\times\times}}
			\imscalar{{h_\times}}{f{h_+}} \rescalar{{h_+}}{\partial_i h_+} \label{eq:H_ti_grad}\\
	H_{ij} &= - \frac{1}{h_{++}} \rescalar{\partial_i h_+}{\partial_j h_+}
		+ \frac{1}{1-\hat{h}^2_{+\times}} \frac{1}{h^2_{++}} \rescalar{h_+}{\partial_i {h_+}} \rescalar{{h_+}}{\partial_j {h_+}}
		+ \frac{1}{1-\hat{h}^2_{+\times}} \frac{1}{h_{++}h_{\times\times}} \rescalar{h_\times}{\partial_i {h_+}} \rescalar{{h_\times}}{\partial_j {h_+}} \nonumber \\
		& - \frac{2 \hat{h}_{+\times}}{1-\hat{h}^2_{+\times}} \frac{1}{h^{3/2}_{++}h^{1/2}_{\times\times}}
		\rescalar{h_\times}{\partial_{(i} {h_+}} \rescalar{{h_+}}{\partial_{j)} {h_+}} \label{eq:H_ij_grad}
%
%
%
\end{align}
\end{widetext}
where we defined $h_{\cdot*} = \rescalar{h_\cdot}{h_*}$.

Such expressions, together with Eq.~\eqref{eq:metric_expression} fully specify the metric.
The gradients $\partial_i h$ of the waveform can be computed with a finite difference scheme or analytically for a number of surrogate waveform models \cite{Khan:2020fso, Schmidt:2020yuu, Thomas:2022rmc, Tissino:2022thn}.

The non precessing limit can be recovered by setting $h_\times = i h_+$ and $h_{+\times} = 0$:
\begin{align}
	H_{tt} &= \frac{1}{h_{++}^{2}} \rescalar{{h_+}}{f{h_+}}^2 - \frac{1}{h_{++}} \rescalar{h_+}{f^2 {h_+}} \label{eq:H_tt_grad_NP} \\
	H_{ti} &= \frac{1}{h_{++}^{2}} \imscalar{h_+}{\partial_i {h_+}} \rescalar{{h_+}}{{h_+}f}
		- \frac{1}{h_{++}} \imscalar{h_+}{f \partial_i{h_+}} \label{eq:H_ti_grad_NP} \\
	H_{ij} &=  \frac{1}{h_{++}^{2}} \Big\{ \rescalar{h_+}{\partial_i {h_+}} \rescalar{{h_+}}{\partial_j {h_+}} +\imscalar{h_+}{\partial_i {h_+}} \imscalar{h_+}{\partial_j {h_+}} \Big\} \nonumber \\
	&- \frac{1}{h_{++}} \rescalar{\partial_i h_+}{\partial_j {h_+}} \label{eq:H_ij_grad_NP} 
\end{align}

\section{Alternative definitions for the metric}\label{app:metric_definition}

\begin{figure}[t]
	\centering
	\includegraphics[scale = .52]{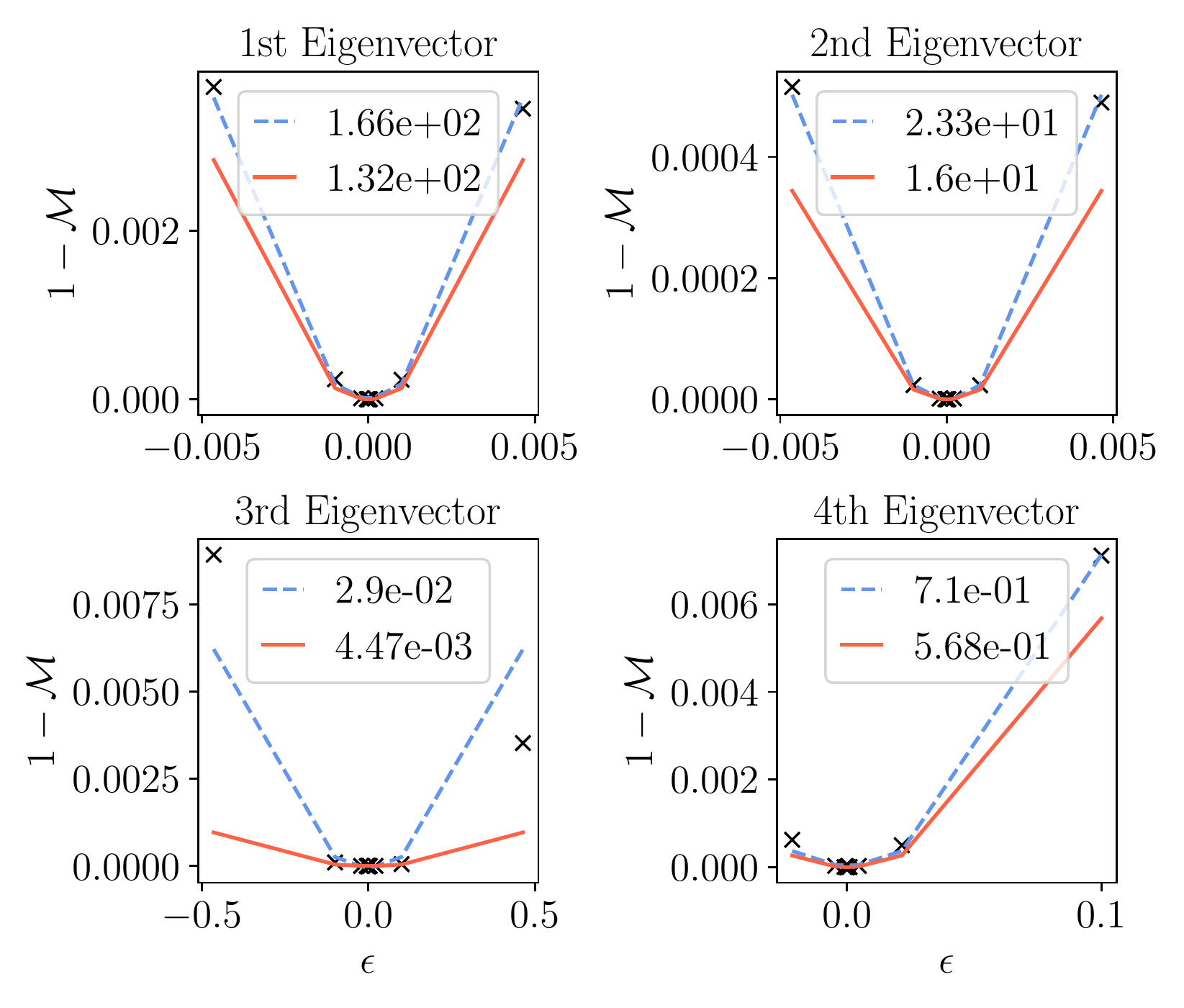}
	\caption{For each eigenvector of the metric, we compute the empirical relation between the mis-match $1-\mathcal{M}$ and the distance $\epsilon$ of points along the eigenvector direction. The solid line shows the relation predicted by the metric, while the dashed line shows a parabolic fit. In the legend are reported the quadratic coefficients of both lines.}
	\label{fig:parabolae}
\end{figure}

Throughout this paper, we identified the metric with the Hessian of the overlap (see Eq.~\eqref{eq:metric_expression}). While this is widely used in the literature \cite{owen_metric, Messenger:2008ta} and has been proven to provide reliable template banks, it still has some undesirable properties.
To show this, we compute the metric at point $\theta_0 = (\SI{20}{\mathrm{M_\odot}}, 3., 0.7, 1.8)$ of manifold \texttt{Mq\_s1xz}, described in Sec.~\ref{sec:flow_validation}, and we compute its eigenvalues $\alpha^{(i)}$ and eigenvectors $v^{(i)}$ . We then compute the match $\mathcal{M}^{(i)}_\epsilon$ between $\theta_0$ and the point $\theta^{(i)}_\epsilon = \theta_0 + \epsilon v^{(i)}$, located at a distance $\epsilon$ along i-th eigenvector.
Finally, we  compute the coefficient $\alpha$ of the Taylor expansion $1 - \mathcal{M}^{(i)}_\epsilon = \alpha  \epsilon^2$.
$\alpha$ corresponds to the i-th eigenvalue and in principle, it should be close to its value.

In Fig.~\ref{fig:parabolae}, we plot the fitted relation between $1 - \mathcal{M}$ and $\epsilon$ for each eigenvector, as well as the one computed with the metric. In the legend we report the $\alpha$ coefficient (dashed blue line) and the eigenvalue of the metric (solid orange line).
The striking feature we note in Fig.~\ref{fig:parabolae} is that the eigenvalue is consistently smaller than the fitted $\alpha$ coefficient, sometimes by an order of magnitude.
This means that the Hessian, which is computed for $\epsilon\rightarrow 0$, is not able to extrapolate the behaviour of $1 - \mathcal{M}(\epsilon)$ even at modestly large value of $\epsilon$: the metric approximation to the match loses its predictivity as a measure of distance.
The problem becomes more severe in high-dimensional manifolds.
On the other hand, since the banks generated with the Hessian metric show nice coverage, one may argue that the {\it volume} estimate provided by the Hessian is still accurate enough for our purposes.

As a way out, we could redefine the matrix $M_{ij}(\theta)$ to a more suitable expression, departing from the Hessian.
The goodness of the metric expression may depend on the application and on the range of validity of the approximation.
The tensor field $M_{ij}(\theta)$ can be computed through an optimization problem, where we minimize the discrepancy between the two quantities in Eq.~\eqref{eq:metric_definition}, encoded into a {\it loss function}.
The loss function depends on the values of the matrix elements $M^\prime_{ij}$:
\begin{equation} \label{eq:loss_function}
	\mathcal{L}_\theta(M^\prime_{ij}) = \hspace{-4em} \int\limits_{\hspace{3em}\{d(\theta,\theta^\prime) < d_\mathrm{target}\}} \hspace{-3.8em}
		\dvol{\theta^\prime}{D}  \left[ 1 - \mathcal{M}(\theta,\theta^\prime) - M^\prime_{ij} \Delta\theta_i \Delta\theta_j \right]^2
\end{equation}
where the integration extends on a D-ball with radius $d_\mathrm{target}$ centered around $\theta$ and $d_\mathrm{target}$ is a tunable parameter, which controls the validity of the approximation.

At any given point $\theta$, the components $M_{ij}(\theta)$ of the metric are selected by minimizing the above loss:
\begin{equation} \label{eq:metric_optmization}
	M_{ij}(\theta) = \argmin_{M^\prime_{ij}}  \mathcal{L}_\theta(M^\prime_{ij}).
\end{equation}
Although the minimization can be tackled with standard techniques, it requires many evaluations of Eq.~\eqref{eq:distance} and the ability to sample from a ``complex" set such as ${\{d(\theta,\theta^\prime) < d_\mathrm{target}\}}$.

While in most cases this may prove unfeasible, future work could solve the problem in Eq.~\eqref{eq:metric_optmization} at a manageable cost. This may be beneficial to many data analysis applications, such as template placement and Fisher information matrix studies.
A number of alternative metric expressions, coming from different heuristic optimization strategies, are already available in \texttt{mbank}, although not fully validated.

\section{Computing the volume of the parameter space}\label{app:parameter_space_volume}

As the number of templates is proportional to the volume of the parameter space \cite{owen_metric}, it can be useful to estimate the volume of the parameter space. This can be useful to forecast the size of a template bank.
The volume can be easily estimated by {\it importance sampling} and, as the normalizing flow reproduces the volume element, it is a convenient distribution to generate samples.

The volume of the parameter space $\mathcal{B}_D$ is defined as:
\begin{align}
	\mathcal{V} & = \int_\mathcal{\mathcal{B}_D} \dvol{\theta}{D} \; \sqrt{\det M(\theta)} \\
				& = \int_\mathcal{\mathcal{S}_\text{flow}} \dvol{\theta}{D} \; \sqrt{\det M(\theta)} \;\; \mathcal{I}_{\mathcal{B}_D}(\theta) \label{eq:volume_indicator_function}
\end{align}
where in the last equality we compute the integral on the support of the flow $\mathcal{S}_\text{flow} \supseteq \mathcal{B}_D$ and we introduced the indicator function $\mathcal{I}_{\mathcal{B}_D}$ which is non-zero only on the manifold $\mathcal{B}_D$.

Eq.~\eqref{eq:volume_indicator_function} can be numerically evaluated by importance sampling:
\begin{align}\label{eq:vol_IS}
	\mathcal{V} \simeq \frac{1}{N} \sum_i \; \frac{\sqrt{\det M(\theta_i)}}{p^\text{flow}(\theta_i)} \; \mathcal{I}_{\mathcal{B}_D}(\theta_i)
\end{align}
with  $\theta_i \sim p^\text{flow}$.
The normalizing flow ensures a low variance in the volume estimation.

Eq.~\eqref{eq:vol_IS} involves several metric evaluations, which has some computational cost.
To further reduce the computational cost, we can use the fact that, after the training procedure, the flow approximates the volume element as follows:
\begin{equation}
	\log p^\text{flow} - \log\sqrt{|M|} + C \simeq 0
\end{equation}
where $C$ is the trainable constant appearing in Eq.~\eqref{eq:loss_mse}.
Hence we can replace $\frac{\sqrt{\det M(\theta_i)}}{p^\text{flow}(\theta_i)}$ in Eq.~\eqref{eq:vol_IS} simply with $e^C$.
The volume estimation is then reduced to computing the fraction of the volume of $\mathcal{S}_\text{flow}$ covered by $\mathcal{B}_D$:
\begin{equation}
	\mathcal{V} \simeq e^C \; \frac{1}{N}  \sum_i \; \; \mathcal{I}_{\mathcal{B}_D}(\theta_i)
\end{equation}
where again $\theta_i \sim p^\text{flow}$.
The goodness of such approximation is closely related to the flow performance, as studied in Sec.~\ref{sec:flow_validation} (see also Fig.~\ref{fig:flow_validation}).

Once an estimation of the volume is available, the number of templates can be obtained by noting \cite{owen_metric} that in a lattice, given a minimal match $MM$, the average spacing $d$ between template is:
\begin{equation}
	d(MM) = 2 \, \sqrt{\frac{1-MM}{D}}
\end{equation}
Hence, roughly speaking, the number of templates $N$ needed to cover the volume $\mathcal{V}$ is given by:
\begin{equation} \label{eq:N_templates}
	N = \frac{\mathcal{V}}{d(MM)^D}
\end{equation}

\begin{figure*}
	\centering
	\includegraphics[scale = 1.]{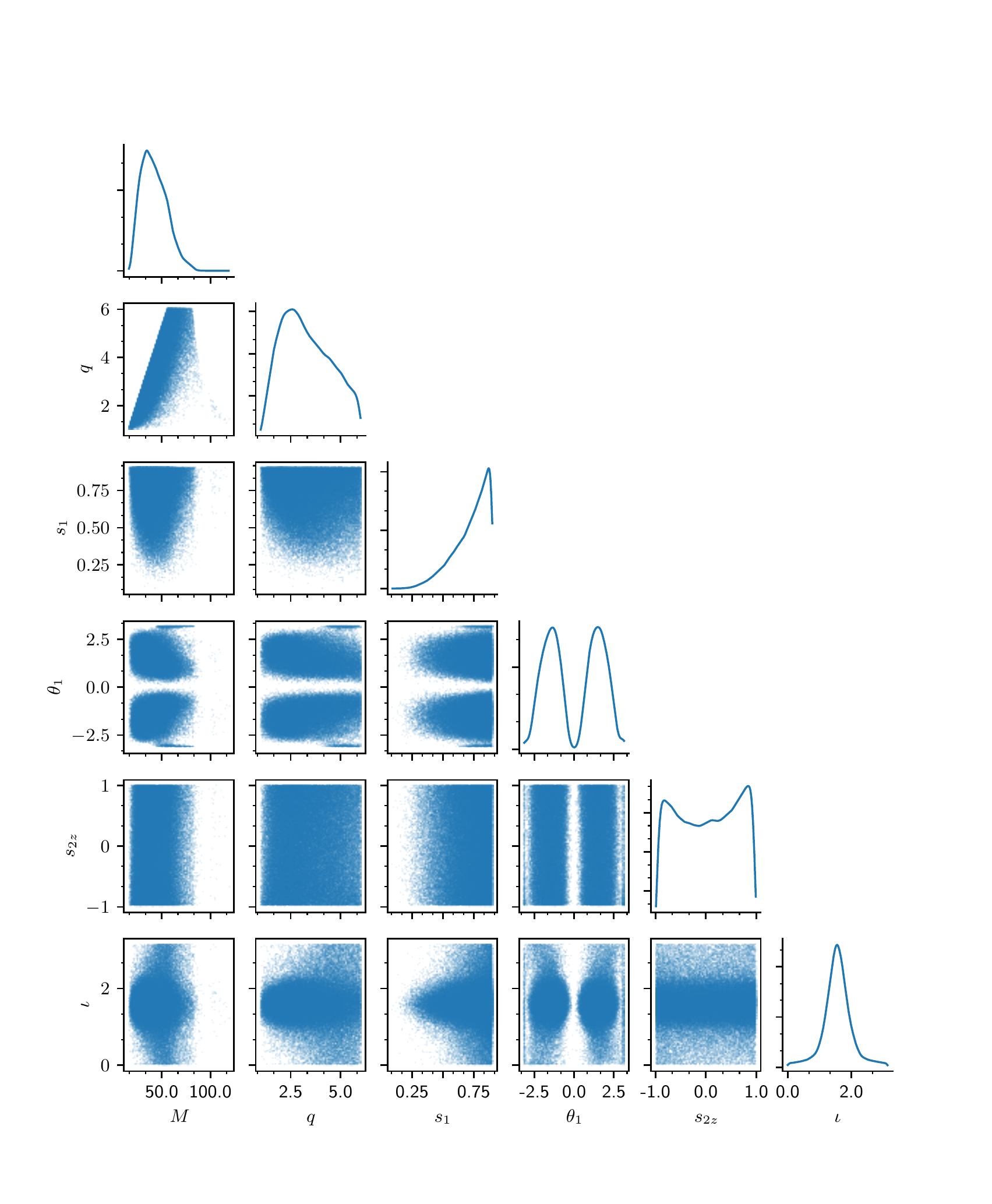}
	\caption{Corner plot with the templates of the precessing bank described in Sec.~\ref{sec:precessing_bank}. Along the diagonals, we show the histogram of the template number as a function of each coordinate.}
	\label{fig:corner_precessing}
\end{figure*}

\begin{figure*}
	\centering
	\includegraphics[scale = 1.]{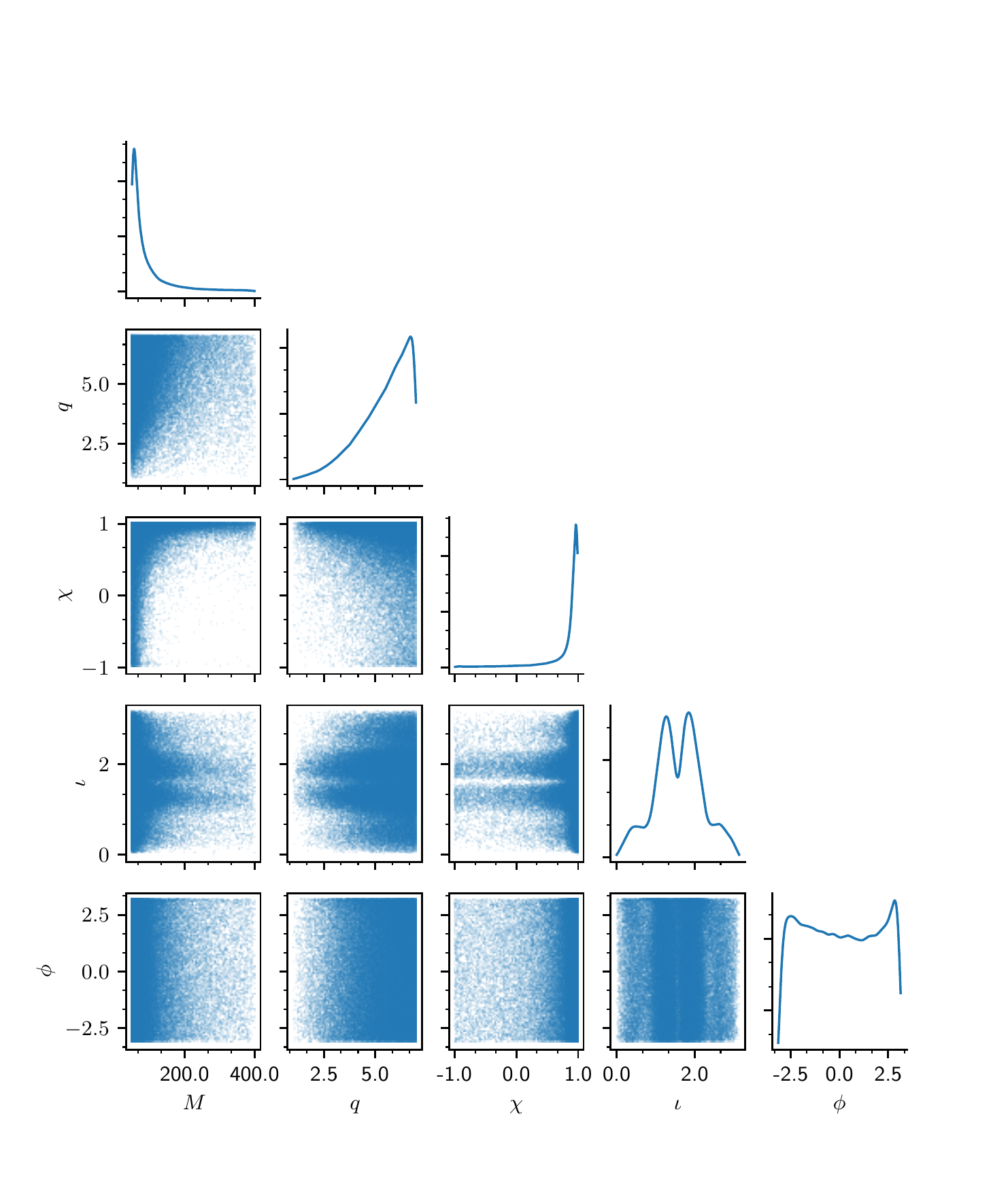}
	\caption{Corner plot with the templates of the aligned-spin HM bank described in Sec.~\ref{sec:HM_spinning_bank}. Along the diagonals, we show the histogram of the template number as a function of each coordinate.}
	\label{fig:corner_HM_spinning}
\end{figure*}

	\bibliography{biblio.bib}
	\bibliographystyle{ieeetr}

\end{document}